\documentstyle[preprint,pra,aps,tighten,eqsecnum,psfig]{revtex}
\begin{document}
\preprint{}
\draft

\title{Classical and quantum measurements of position}
\author{Carlo Presilla,\cite{CARLO}}
\address{Dipartimento di Fisica, Universit\`a di Roma ``La Sapienza,''\\
Piazzale A. Moro 2, Roma, Italy 00185}
\author{Roberto Onofrio,\cite{ROBERTO}}
\address{Dipartimento di Fisica ``G. Galilei,'' Universit\`a di Padova,\\
Via Marzolo 8, Padova, Italy 35131}
\author{Marco Patriarca\cite{MARCO}}
\address{Dipartimento di Chimica, Universit\`a di Roma ``La Sapienza'',\\
Piazzale A. Moro 2, Roma, Italy 00185}
\maketitle

\begin{abstract}
We study the dynamics of classical and quantum systems undergoing a 
continuous measurement of position by schematizing the measurement apparatus 
with an infinite set of harmonic oscillators at finite temperature linearly 
coupled to the measured system.
Selective and nonselective measurement processes are then introduced according
to a selection of or an average over all possible initial configurations 
of the measurement apparatus.
At quantum level, the selective processes are described by a nonlinear 
stochastic Schr\"odinger equation whose solutions evolve into properly defined
coherent states in the case of linear systems.
For arbitrary measured systems, classical behavior is always recovered in 
the macroscopic limit.
\end{abstract}

\section{Introduction}

A fundamental problem in quantum mechanics is the relationship 
between the states in the Hilbert space of a quantum system and the 
states in the phase space of the corresponding classical system. 
This is particularly evident in the case of a superposition of states 
which are individually mapped in the macroscopic limit, formally $\hbar \to 0$,
into distinguishable classical states \cite{SCHROEDINGER,EINSTEIN}.
 
One important step toward the solution of this problem has been made by 
recognizing that a system is never completely isolated by the external world. 
It has been argued that an external environment can, after a transient whose
duration presumably depends on the coupling strength, drive the totality of 
the admissible states of the Hilbert space into those having macroscopic 
limit \cite{JOOSZEH,ZUREK,CINI}. 

Among all the conceivable situations which require the interaction with 
an environment, a peculiar role is played by the measurement processes.
Indeed, whenever any physical property of a system is investigated, an
unavoidable coupling with the degrees of freedom of the measurement apparatus
must be invoked.
Taking into account these external degrees of freedom naturally provides 
a generalization of the von Neumann postulate \cite{VONNEUMANN} to continuous
measurements \cite{WHEE,BELAVKIN,PREONTA}. 

There exists a basic difference between a general environment and one
schematizing a measurement process.
According to the Copenhagen interpretation, classical behavior of the 
measurement apparatus has to be assumed before the information 
is registered by the observer \cite{CINI,BOHR,HAAKE}.
This requirement, although establishing a classical connection between 
observer and meter, does not imply a classical behavior of the observed system
since its interaction with the meter fully preserves quantum features. 
This aspect may be emphasized by comparing the coupling of the same
measurement apparatus, classically controlled by the observer,
to classical or quantum systems.

This paper is devoted to establishing the relationship between the dynamics of 
classical and quantum systems under the influence of a measurement process.
We model the effect of the measurement by allowing the measured system
to linearly interact with an infinite set of harmonic oscillators. 
The interaction occurs in the configuration space thus representing a direct
measurement of position. 
Our approach shouldn't be understood as the modelling of a particular
device, e.g., a cloud chamber in which the oscillators
represent molecules distributed in a medium \cite{WANHARRISON}.
Our oscillators are the modes of an interaction field which 
reproduces the main features expected in a measurement process, 
e.g., the wavefunction collapse, without {\it ad hoc} rules. 
Differently from other abstract models \cite{FRAASSEN},
our approach has, however, a certain degree of realism since allows us 
to operatively define the reading of the measurement results.

The oscillators representing the measurement device 
are chosen at thermal equilibrium with temperature $T$ and
continuously distributed in frequency with a proper density.
These conditions ensure that, for a chosen initial configuration of the
oscillators,  the classical measured system is described by a Markovian 
Langevin equation with white noise and a certain relaxation time $\gamma^{-1}$.
The constant $\gamma$, which fixes the magnitude of the density of oscillators,
represents the strength of the measurement process.
In the quantum case, a further characteristic time of the measurement 
apparatus, $\hbar/k_B T$, arises.
The requirement of classical behavior of the meter with respect to the observer
implies a high temperature condition to hold so that this thermal 
fluctuation time must be much shorter than the relaxation one.
As a consequence, the measured quantum system is described by a stochastic 
nonlinear Schr\"odinger equation  
\cite{DIOSI88A,DIOSI88B,GISINPERCIVAL,PERCIVAL}.

Details of the derivation of the stochastic equations describing 
classical and quantum systems during a measurement process are given in
Sections II and III, respectively.
We call these processes selective since they correspond to a single measurement
act with initial conditions of the meter selected among those compatible with 
the assumed thermal equilibrium.
Alternatively, one can also consider nonselective measurement processes
corresponding to an average over all initial configurations of the meter.
In this case, the dynamics of classical and quantum measured systems is
described by a Fokker-Plank equation and a trace-preserving positive 
master equation, respectively. 
The experimental results obtained by repeating a measurement on a 
system always in the same initial state or performing the same measurement on 
an ensemble of equally prepared independent systems 
is directly comparable with the solution of these nonselective equations.
Equivalent results are predicted by averaging the solutions of the selective
equations, classical or quantum, over the corresponding noise realizations. 

In Section IV we show how to infer the measurement results by the reading 
of an appropriate pointer. 
Since the oscillators representing the meter are classical
with respect to the observer, the pointer can be defined in terms
of the coordinates of these oscillators which, in turn, reflect the status 
of the measured system.

As for closed quantum systems, also in presence of
a measurement process there exists a class of states, namely
the coherent ones, which admits, in a proper sense, the $\hbar \to 0$ limit.
These coherent states, explicitly built in Section V, are Gaussian localized 
states in the co-moving frame of a measured linear system 
\cite{DIOSI88B,GRW,HALLIWELL95B}.
In Section VI we show that for linear systems 
the solutions of the stochastic Schr\"odinger equation converge to a 
coherent state localized around a point in the phase space which moves 
according to a Langevin-like equation.
This equation reduces to the classical one for $\hbar \to 0$.
The convergence into this coherent state occurs in a timescale 
${(\hbar/\gamma k_B T)}^{1/2}$, the geometric mean of the two characteristic
times associated to the measurement apparatus.
This time diverges for an unmeasured system ($\gamma \to 0$)  
and vanishes in the macroscopic limit.

In the case of nonlinear systems, the phase-space localization through 
convergence into a coherent state becomes the leading dynamical process 
as $\hbar \to 0$.
This is sufficient to demonstrate that classical behavior is always 
recovered in the macroscopic limit, avoiding any paradoxical quantum feature.

\section{Classical systems} 
Let us consider a system described by the Hamiltonian
\begin{equation} 
H(p,q,t) = {p^2 \over {2m}}+V(q,t)
\label{H}
\end{equation}
and suppose that we want to measure its position $q$ 
($q \in \bbox{R}$, for simplicity).
We schematize the measurement process by the interaction of the system
(\ref{H}) with a set of particles of mass $M$ and canonical coordinates 
$(P_n, Q_n)$ via a harmonic potential which, for evident physical reasons,
must depend on the relative distances $Q_n-q$.
The Hamiltonian for the total system is then taken as
\begin{equation} 
H_{tot} = H(p,q,t) + H_m(P,Q-q), 
\label{HTOT}
\end{equation}
where
\begin{equation}
H_m(P,Q-q) = \sum_n \left[ {P_n^2 \over {2M}}+
{M \omega_n^2 \over 2} (Q_n-q)^2 \right]
\label{HM}
\end{equation}
represents the measurement apparatus linearly coupled to the measured system.
Here and in the following, $Q$ and $P$ are a shortening for the whole set 
$\{Q_n\}$ and $\{P_n\}$. 
Note that the Hamiltonian (\ref{HM}) can be interpreted as that of a 
set of harmonic oscillators with equilibrium positions $Q_n=q$.
Our model is thus reduced to the exactly solvable problem of a system 
interacting with a bath of harmonic oscillators \cite{CALDLEGG} but with the 
difference that the interaction potential is invariant under space 
translations.    
The importance of this invariance in avoiding the appearance of
infinite renormalization potentials has been recently underlined 
in \cite{PATRIARCA} in the framework of the classical/quantum Brownian motion.
We stress that the long range nature of the quadratic potential in
(\ref{HM}) prevents us from considering a situation in which the measured 
system does not interact with the meter so that our model cannot 
take into account the switching-on of the measurement.

At classical level, the Newton equation for each harmonic oscillator
can be explicitly solved in terms of the function $q(t)$ and 
the values of the coordinates $Q_n'=Q_n(t')$, $P_n'=P_n(t')$,
and $q'=q(t')$ at the initial time $t'$
\begin{eqnarray}
Q_n(t) &=& (Q_n'-q') \cos[\omega_n (t-t')] + 
\frac{P_n'}{M \omega_n} \sin[\omega_n (t-t')] 
\nonumber \\ &&
+q(t)-\int_{t'}^t ds \cos[\omega_n (t-s)]~\dot{q}(s).
\label{CQSOL}
\end{eqnarray}
We will also assume $p(t')=p'$.
When this solution is inserted in the Newton equation for the measured 
system, the following equation for $q(t)$ is obtained
\begin{eqnarray}
m \ddot{q}(t) 
+ \int_{t'}^{t} ds~ \Gamma(t-s)~ \dot{q}(s) 
+ \partial_q V(q(t),t) = \Pi(t), 
\label{LANGE}
\end{eqnarray}
where 
\begin{equation}
\Gamma(t-s)  = \sum_{n} M \omega_n^2 \cos[\omega_n (t-s)]
\label{GAMMA}
\end{equation}
and 
\begin{eqnarray}
\Pi(t) = 
\sum_n M \omega_n^2
\left\{ (Q_n'-q') \cos[\omega_n (t-t')] + 
\frac{P_n'}{M \omega_n} \sin[\omega_n (t-t')] \right\}.
\label{PI}
\end{eqnarray}
In the limit of an infinite number of oscillators,
if the corresponding initial conditions $Q'$ and $P'$ are a realization
of a stochastic process in the phase space, $\Pi(t)$ is a realization
of a stochastic process in time.
If, as we will suppose, the oscillators are at thermal equilibrium with
temperature $T$,
the initial conditions to be considered are typical realizations of the 
stochastic process corresponding to the equilibrium Gibbs measure \cite{JP}.
In this case, the following statistical properties for $\Pi(t)$ hold
\begin{equation}
\overline{\Pi(t)} =0, ~~~~~ 
\overline{\Pi(t) \Pi(s)} = k_B T~\Gamma(t-s),
\label{FLUDIS}
\end{equation}
where
\begin{equation}
\overline{\cdots} = 
{ \int dP' dQ'~ \cdots ~ e^{- H_m(P',Q'-q') / k_BT} \over
  \int dP' dQ'~ e^{- H_m(P',Q'-q') / k_BT} }.
\label{CAVEDEF}
\end{equation}
Note that the initial conditions corresponding to the definition 
(\ref{CAVEDEF}) respect the space translation symmetry of the Hamiltonian
$H_m$ thus implying, in agreement with the long range nature of the 
quadratic potential, a correlation between the coordinates $Q'$ and $q'$.

The friction term in the stochastic differential equation (\ref{LANGE}) 
contains memory effects which are an unessential complication in our context.
A Markovian evolution can be obtained by choosing an appropriate continuous 
distribution of the frequencies $\{ \omega_n \}$.
For a frequency density 
\begin{equation}
{dN \over d\omega} = {2 m \gamma \over \pi M \omega^2} 
\theta(\Omega-\omega),
\label{DENSITY}
\end{equation}
where $\theta(x)$ is 1 for $x>0$ and $0$ otherwise, 
we get
\begin{equation}
\Gamma(t-s)=
\int_0^{\infty} d\omega \frac{dN}{d \omega}{M \omega^2}
\cos [\omega(t-s)] = 2 m \gamma {\sin[\Omega(t-s)]\over \pi (t-s)}
\simeq 2 m \gamma \delta(t-s) .
\end{equation}
The approximation holds for $t-s \gg \Omega^{-1}$.
For $\Omega^{-1} \ll \tau$, where $\tau$ is the fastest time scale at 
which the measured system evolves, $\Pi(t)$ can be approximated with a 
white noise and Eq. (\ref{LANGE}) rewritten as
\begin{eqnarray}
dp(t) &=& - \left[ \gamma p(t) + \partial_q V(q(t),t) \right] dt + 
\sqrt{ 2 m \gamma k_B T} dw(t)
\label{LANGE1} \\
dq(t) &=& {p(t) \over m} dt.
\label{LANGE2}
\end{eqnarray}
Here, we introduced the Wiener process 
$dw(t) = (2 m \gamma k_B T)^{-1/2} \Pi(t)dt$  
having zero average, $\overline{dw(t)} =0$, and standard scaling, 
$\overline{dw(t) dw(t)} =  dt$.
Note that the friction coefficient $\gamma$ and the temperature $T$ 
completely define the fluctuation and dissipation phenomena induced by 
the interaction with the measurement apparatus.

Equations (\ref{LANGE1}) and (\ref{LANGE2}) describe the evolution of the 
system during a selective measurement, i.e., a measurement in which a 
realization of the stochastic process $p(t)$, $q(t)$ is selected according 
to certain initial conditions of the measurement apparatus.
Alternatively, one can consider a nonselective measurement corresponding
to an average over all possible realizations of the stochastic process.
In this case, the measured system is described by a probability density
$W(p,q,t)$ which is determined by the Fokker-Plank equation \cite{KUBO}
associated to (\ref{LANGE1}) and (\ref{LANGE2}) 
\begin{equation}
\partial_t W(p,q,t) = 
\left[ - {p \over m} \partial_q + \partial_q V(q,t) \partial_p 
+ \partial_p \left( \gamma p + m \gamma k_BT \partial_p \right)
\right] W(p,q,t)
\label{FOKKERPLANK}
\end{equation}
with initial conditions $W(p,q,t')= \delta(p-p') \delta(q-q')$.
The probability density $W(p,q,t)$ allows us to directly evaluate 
averages of any function of $p(t)$ and $q(t)$.
In particular, for the position we have the average value
\begin{equation}
\overline{q(t)} = \int ~ dpdq ~ W(p,q,t) ~ q
\label{CAVE}
\end{equation}
with a variance 
\begin{equation}
\Delta q^2(t)= \int ~ dpdq ~ W(p,q,t)~\left[ q-\overline{q(t)} \right]^2
= \overline{q(t)^2} - \overline{q(t)}^2.
\label{CVAR}
\end{equation}
For $\gamma \to 0$ the effect of the measurement vanishes.
In this case, Eq. (\ref{FOKKERPLANK}) becomes the Liouville equation for
the isolated system (\ref{H}), the average value (\ref{CAVE}) gives the
corresponding time dependent solution, and the variance (\ref{CVAR}) vanishes.

\section{Quantum systems} 
At quantum level, the measured system is conveniently described by a reduced 
density matrix obtained by tracing out the coordinates of the measurement 
apparatus in the total density matrix 
\begin{equation}
\varrho(q_1, q_2, t) = 
\int dQ ~ \varrho_{tot}(q_1, Q, q_2, Q, t).
\end{equation}
We assume that at the initial time $t'$ the oscillators of the measurement 
apparatus are at thermal equilibrium with temperature $T$ and the 
total density matrix is factorized as 
\begin{equation}
\varrho_{tot}(q_1',Q_1',q_2',Q_2',t')=\varrho(q_1',q_2',t') 
\varrho_m (\delta Q_1',\delta Q_2',t'),
\label{INIRHO}
\end{equation}
where
\begin{eqnarray}
\varrho_{m} (\delta Q_1',\delta Q_2',t') =&&
\prod_{ n} 
\sqrt{ \frac{M \omega_{n}}{\pi \hbar }
\tanh \left( {\hbar \omega_{n} \over 2 k_B T} \right) } 
\nonumber \\
&&~ \times \exp \Bigg[ 
- \frac{M \omega_{n}} {2 \hbar }
\Bigg( 
{ \delta Q_{1n}'^2 + \delta Q_{2 n}'^2 
\over \tanh(\hbar \omega_{ n}/k_BT) }
- { 2 \delta Q_{1 n}' \delta Q_{2 n}' 
\over \sinh(\hbar \omega_{ n}/k_BT) }
\Bigg) \Bigg]
\label{RHOENV}
\end{eqnarray}
with $\delta Q_1' = Q_1' - Q'_{eq}$ and $\delta Q_2' = Q_2' - Q'_{eq}$.
The requirement of translational and time reversal invariance 
implies \cite{PATRIARCA}
\begin{equation}
Q'_{eq} = {q_1' + q_2' \over 2}.
\label{QEQU}
\end{equation}
Analogously to the classical case, this choice corresponds to an 
initial condition (\ref{INIRHO}) factorized but correlated.
 
At a later time $t$, the reduced density matrix is obtained through a 
Green function
\begin{equation}
\varrho(q_1, q_2, t) = 
\int dq_1' dq_2'  
~G( q_1,q_2,t;q_1',q_2',t')~
\varrho(q_1',q_2',t')  
\label{INTEGRALEVOL}
\end{equation}
whose path-integral representation 
\begin{eqnarray}
G(q_1,q_2,t;q_1',q_2',t') &=& 
\int d[p_1] d[q_1]_{q_1',t'}^{q_1,t} 
\int d[p_2] d[q_2]_{q_2',t'}^{q_2,t}
\nonumber \\ && \times 
\exp \left\{ \frac{i}{\hbar} S[p_1,q_1] - \frac{i}{\hbar} S[p_2,q_2] 
-Z[p_1,q_1,p_2,q_2] \right\}
\label{GREENF}
\end{eqnarray}
is the free evolution of the measured system modified by the influence 
functional
\begin{eqnarray}
\exp \{-Z[p_1,q_1,p_2,q_2] \} &=&   
\int dQ \int dQ_1' dQ_2'
\int d[P_1] d[Q_1]_{Q_1',t'}^{Q,t}
\int d[P_2] d[Q_2]_{Q_2',t'}^{Q,t} \nonumber \\&\times&
\exp \left\{ 
  \frac{i}{\hbar} S_m[P_1,Q_1-q_1] 
- \frac{i}{\hbar} S_m[P_2,Q_2-q_2] \right\}
\varrho_{m} (\delta Q_1',\delta Q_2',t').
\label{INF}
\end{eqnarray}
Here, $S[p,q]$ and $S_m[P,Q-q]$ are the classical actions 
corresponding to the Hamiltonians $H$ and $H_m$, respectively
\begin{eqnarray}
S[p,q]=\int_{t'}^t ds [p(s)\dot{q}(s)-H(p,q,s)]
\end{eqnarray}
\begin{eqnarray}
S_m[P,Q-q]=\int_{t'}^t ds [P(s)\dot{Q}(s)-H_m(P,Q-q,s)].
\end{eqnarray}
The functional measure with boundary conditions $q(t')=q'$ and $q(t)=q$  
is obtained by slicing the interval $[t',t]$ at times $t^{(n)}=t'+(t-t')n/N$,
$n=1,\ldots,N$, and taking the $N \to \infty$ limit.
All time integrals can be approximated with sums
\begin{eqnarray}
\int_{t'}^{t} ds f(p,q,s) = 
\sum_{n=1}^N \int_{t^{(n-1)}}^{t^{(n)}} ds f(p,q,s) \simeq
{t-t' \over N} \sum_{n=1}^N f\left( p^{(n)},q^{(n)},t^{(n)} \right),
\end{eqnarray}
where $p^{(n)}=p(t^{(n)})$ and $q^{(n)}=q(t^{(n)})$, and 
\begin{eqnarray}
d[p] d[q]_{q',t'}^{q,t} =  \lim_{N \to \infty}
\prod_{n=1}^{N} {dp^{(n)} \over 2 \pi \hbar} 
\prod_{n=1}^{N-1} dq^{(n)}.
\label{DPDQ}
\end{eqnarray}
Analogous relations hold for $d[P] d[Q]_{Q',t'}^{Q,t}$.

The influence functional (\ref{INF}) contains only Gaussian integrals and
can be evaluated exactly. The result is \cite{FV}
\begin{eqnarray}
Z[p_1,q_1,p_2,q_2] &=&  
\int_{t'}^{t} ds \int_{t'}^{s} du ~
\left[ q_1(s) - q_2(s) \right] \nonumber \\ &&\times
\left\{ \Lambda(s-u)  \left[ q_1(u) - q_2(u) \right]  
+ {i \over 2 m \hbar} \Gamma(s-u) \left[ p_1(u) + p_2(u) \right]
\right\},
\label{GFI}
\end{eqnarray}
where 
\begin{equation}
\Lambda(s-u) = \frac{M}{2 \hbar} \sum_n \omega_{n}^3
\coth \left( \frac{\hbar \omega_{n}}{2 k_B T} \right) 
\cos[\omega_{n} (s-u)]
\label{FLU1}
\end{equation}
and $\Gamma(s-u)$, defined by Eq. (\ref{GAMMA}), 
are named fluctuation and dissipation kernels, respectively \cite{CALDLEGG}.
The double time integral which appears in Eq. (\ref{GFI}) is responsible
for memory effects which break the semi-group property of the evolution of the
measured system, i.e., 
\begin{equation}
G(q_1,q_2,t;q'_1,q'_2,t') = \int dq''_1 dq''_2~ 
G(q_1,q_2,t;q''_1,q''_2,t'')~G(q''_1,q''_2,t'';q'_1,q'_2,t')
\label{SEMIGROUP}
\end{equation}
for $t'<t''<t$. 
Two are the sources of this non-Markovian behavior.
In the dissipation kernel $\Gamma$, the origin of memory effects is classical 
and Markovian behavior is obtained if one assumes the frequency distribution 
(\ref{DENSITY}) with $\Omega^{-1} \ll \tau$.
In the fluctuation kernel $\Lambda$, the assumption of these conditions
does not remove the memory effects due to the quantum behavior of the
oscillators in the thermal bath.
However, in the framework of a theory of measurement processes, 
the measurement apparatus, which is an interface between the observer 
(classical) and the measured system (quantum), must have classical 
behavior with respect to the former in order to avoid paradoxical features
\cite{CINI}.
In the present model, the nature of the interaction between the measurement 
apparatus and the external environment is fixed by the ratios 
$\hbar \omega_n / k_BT$ so that the requirement of classical behavior
of the measurement apparatus with respect to the observer 
imposes the condition $k_B T \gg \hbar \Omega$.

For $\hbar/k_BT \ll \Omega^{-1} \ll \tau$ and frequency distribution 
(\ref{DENSITY}), Eq. (\ref{GFI}) can be approximated (see Appendix A) with
\begin{eqnarray}
Z[p_1,q_1,p_2,q_2] =   
\int_{t'}^{t} ds \bigg\{ {m \gamma k_BT \over \hbar^2}
\left[ q_1(s) - q_2(s) \right]^2
+ {i \gamma \over 2 \hbar}  \left[ q_1(s) - q_2(s) \right]  
\left[ p_1(s) + p_2(s) \right] \bigg\}. 
\label{CLIF}
\end{eqnarray}
A differential equation for the reduced density matrix operator can be then 
derived with standard methods. 
According to Eq. (\ref{CLIF}) and using the path-integral representation 
of $G$,  we have
\begin{eqnarray}
\varrho(q_1,q_2,t+dt) &=& \int dq_1' dq_2'  
~G( q_1,q_2,t+dt;q_1',q_2',t)~\varrho(q_1',q_2',t)  
\nonumber \\ &=&
\int {dp_1 \over 2 \pi \hbar} dq_1'
\int {dp_2 \over 2 \pi \hbar} dq_2' 
~\exp\bigg\{  
 {i\over\hbar} p_1 \left(q_1-q_1'\right) 
-{i\over\hbar} p_2 \left(q_2-q_2'\right)
\nonumber \\&& 
+ \bigg[ -{i\over\hbar} H\left(p_1,q_1,t\right)
+{i\over\hbar} H\left(p_2,q_2,t\right)
- {m \gamma k_BT \over \hbar^2} \left(q_1-q_2\right)^2
\nonumber \\&& 
- {i \gamma \over 2 \hbar} \left(q_1-q_2\right)
\left(p_1+p_2 \right) \bigg] dt \bigg\} 
\varrho\left(q_1',q_2',t\right).
\end{eqnarray}
By using the identity 
$\langle q | p \rangle = (2\pi\hbar)^{-1/2} \exp(ipq/\hbar)$
and expanding the exponential containing the infinitesimal time $dt$, 
the above equation can be cast in the form  
\begin{equation}
\langle q_1 | \hat{\varrho}(t+dt) | q_2 \rangle = 
\langle q_1 | \hat{\varrho}(t) | q_2 \rangle + 
\langle q_1 | {d\over dt}\hat{\varrho}(t) | q_2 \rangle,
\end{equation}
where
\begin{eqnarray}
{d\over dt} \hat{\varrho}(t) = 
-{i\over\hbar} \left[ \hat{H}(\hat{p},\hat{q},t),\hat{\varrho}(t) \right]
- {m \gamma k_BT \over \hbar^2}
\left[ \hat{q}, \left[ \hat{q},\hat{\varrho}(t) \right] \right]
- {i \gamma \over 2 \hbar}
\left[ \hat{q}, \left\{ \hat{p},\hat{\varrho}(t) \right\} \right].
\label{CLMEQ}
\end{eqnarray}
Note that $d[\mbox{Tr} \hat{\varrho}(t)]/dt = 0$ so that we can 
assume $\mbox{Tr} \hat{\varrho}(t) = 1$.

Equation (\ref{CLMEQ}) describes the evolution of the measured 
system with initial conditions of the oscillator system statistically 
distributed according to Eq. (\ref{RHOENV}). 
This is a nonselective measurement process to be compared with the 
classical one described by Eq. (\ref{FOKKERPLANK}).

The quantum-classical correspondence of nonselective measurements can be 
extended to selective processes.
Introducing 
\begin{eqnarray}
A(p(t),q(t)) &=& \sqrt{2m \gamma k_BT \over \hbar^2} ~q(t) + 
i \sqrt{\gamma \over 8 m k_BT} ~p(t) \\
B(p(t)) &=& \sqrt{\gamma \over 8 m k_BT} ~p(t),
\end{eqnarray}
Eq. (\ref{CLIF}) can be rewritten as
\begin{eqnarray}
Z =    
\int_{t'}^{t} ds ~ &\bigg[& {i \gamma \over 2 \hbar} (q_1p_1 -q_2p_2)
+{1\over2}A_1A_1^*+{1\over2}A_2A_2^* 
-A_1A_2^* -{1\over2}(B_1-B_2)^2 \bigg]
\nonumber \\ =    
\int_{t'}^{t} ds ~ &\bigg[& {i \gamma \over 2 \hbar} (q_1p_1 -q_2p_2)
+{1\over2}A_1A_1^*+{1\over2}A_2A_2^* 
-{1\over2}(A_1-a)^2 -{1\over2}(A_2^*-a^*)^2 +aa^*
\nonumber \\ &&
-A_1a^* -A_2^*a
+{1\over2}(A_1-a-A_2^*+a^*)^2 -{1\over2}(B_1-b-B_2+b)^2 \bigg],
\label{ZPREOPEN}
\end{eqnarray}
where $A_i$ and $B_i$ stand for $A(p_i,q_i)$ and $B(p_i)$, $i=1,2$, and
$a(t)$ and $b(t)$ are two arbitrary functions, respectively complex and real.
The coupling between the components 1 and 2 of the system coordinates
given by the last two squares of the exponential can be eliminated
in terms of two functional integrations over white real noises
by means of the identities
\begin{eqnarray}
&&\exp\left[ -{1\over2} \int_{t'}^t ds~ (A_1-a-A_2^*+a^*)^2 
\right] \nonumber \\ &&= \int d[\xi] 
\exp\bigg\{- \int_{t'}^t ds~ \Big[ (A_1-a)^2+(A_2^*-a^*)^2 
-(A_1-a)\xi-(A_2^*-a^*)\xi \Big] 
\bigg\}
\label{OPENA}
\end{eqnarray}
and
\begin{eqnarray}
&&\exp\left[ {1\over2} \int_{t'}^t ds~ (B_1-b-B_2+b)^2 
\right] \nonumber \\ &&= \int d[\eta] 
\exp\bigg\{ \int_{t'}^t ds~ \Big[ (B_1-b)^2+(B_2-b)^2 
+i(B_1-b)\eta+i(B_2-b)\eta \Big] 
\bigg\}.
\label{OPENB}
\end{eqnarray}
Note that the above functional integration measures are Gaussian
\begin{eqnarray}
d[\xi] &=&  \lim_{N \to \infty}
\prod_{n=1}^{N} d\xi^{(n)} \sqrt{t-t' \over 2 \pi N} 
\exp \left( -{ {(t-t') \xi^{(n)}}^2 \over 2N} \right)
\label{XIMEASURE}
\end{eqnarray}
so that 
\begin{equation}
\overline{\xi(t)} = \int d[\xi]~\xi(t) = 0, ~~~~~~~
\overline{\xi(t)\xi(s)} = 
\int d[\xi]~ \xi(t)~\xi(s) = \delta(t-s),
\end{equation}
and analogously for $\eta$.
The two-particle Green function (\ref{GREENF}) can be then rewritten 
in terms of a couple of one-particle Green functions
\begin{equation}
G(q_1,q_2,t;q_1',q_2',t')  
= \int d[\xi] d[\eta]
~G^+_{[\xi \eta]}(q_1,t;q_1',t')  
~G^-_{[\xi \eta]}(q_2,t;q_2',t')^*,  
\end{equation}
where
\begin{eqnarray}
&&G^\pm_{[\xi \eta]}(q,t;q',t') =
\int d[p] d[q]_{q',t'}^{q,t} 
\exp \bigg\{ \frac{i}{\hbar} S[p,q] 
- {i \gamma \over 2 \hbar} \int_{t'}^t ds~pq
\nonumber \\ &&~
+ \int_{t'}^t ds \Big[
-{1\over2}AA^* -{1\over2}(A-a)^2 -{1\over2}aa^* + Aa^* + (A-a)\xi
+(B-b)^2 \pm i (B-b)\eta  \Big] \bigg\}.
\end{eqnarray}
Assuming that the system is initially in a pure state, i.e.,
$\hat{\varrho}(t')=|\psi(t') \rangle \langle \psi(t')|$,
at a later time $t$ the reduced density matrix operator is 
expressed as a functional integral over pure states 
\begin{equation}
\hat{\varrho}(t) = \int d[\xi] d[\eta]~
|\psi^+_{[\xi \eta]}(t) \rangle \langle \psi^-_{[\xi \eta]}(t)|,
\label{RHODECT}
\end{equation}
obtained propagating $|\psi(t') \rangle$ with $G^\pm_{[\xi \eta]}$
\begin{equation}
\langle q | \psi^\pm_{[\xi \eta]}(t) \rangle = 
\int dq'~G^\pm_{[\xi \eta]}(q,t;q',t') ~ 
\langle q' | \psi(t') \rangle.
\label{PSIINTEGRALEVOL}
\end{equation}

An evolution equation for the states $|\psi^\pm_{[\xi \eta]}(t) \rangle$
can be obtained by writing the explicit form of the propagators 
$G^\pm_{[\xi \eta]}$ between the times $t$ and $t+dt$
\begin{eqnarray}
\langle q | \psi^\pm_{[\xi \eta]}(t+dt) \rangle &=& 
\int dq'~G^\pm_{[\xi \eta]}(q,t+dt;q',t)~\langle q' | \psi(t) \rangle
\nonumber \\ &=&
\int {dp \over 2 \pi \hbar} dq'
\exp\bigg\{  
 {i\over\hbar} p\left(q-q'\right) 
+ \bigg[ 
-{i\over\hbar} H\left(p,q,t\right)
- {i \gamma \over 2 \hbar} p q
\nonumber \\ &&
-{1\over2}A(p,q)A(p,q)^* -{1\over2}[A(p,q)-a(t)]^2 
-{1\over2}a(t)a(t)^* + A(p,q)a(t)^* 
\nonumber\\ &&
+ [A(p,q)-a(t)] \xi(t) + [B(p)-b(t)]^2 
\pm i [B(p)-b(t)] \eta(t) \bigg] dt \bigg\}
\nonumber \\ && \times
\langle q' | \psi^\pm_{[\xi \eta]}(t) \rangle.
\end{eqnarray}
By using the identity 
$\langle q | p \rangle = (2\pi\hbar)^{-1/2} \exp(ipq/\hbar)$
and expanding the exponential containing the Wiener processes 
$dw_\xi(t)=\xi(t) dt$ and $dw_\eta(t)=\eta(t) dt$ 
according to the Ito rule \cite{ARNOLD}, 
we get the following stochastic differential equation \cite{NOTA1}
\begin{eqnarray}
d | \psi^\pm_{[\xi \eta]}(t) \rangle &=& 
- {i \over \hbar} \left[ \hat H(\hat p,\hat q,t) +
{\gamma \over 4} (\hat{p}\hat{q}+\hat{q}\hat{p}) \right]
| \psi^\pm_{[\xi \eta]}(t) \rangle~ dt
\nonumber \\ &&
-{1\over2} \left[ \hat{A}^{\dag}\hat{A} + a(t)^*a(t) - 2a(t)^*\hat{A}\right]
| \psi^\pm_{[\xi \eta]}(t) \rangle~ dt
+\left[ \hat{A} - a(t) \right] | \psi^\pm_{[\xi \eta]}(t) \rangle~ dw_\xi(t)
\nonumber \\ &&
+{1\over2} \left[ \hat{B}-b(t) \right]^2 | \psi^\pm_{[\xi \eta]}(t) \rangle~ dt
\pm i\left[ \hat{B}-b(t) \right] |\psi^\pm_{[\xi \eta]}(t) \rangle~ dw_\eta(t).
\label{STOCEQ}
\end{eqnarray}
The normalization condition for the reduced density matrix operator 
\begin{equation}
\mbox{Tr} \hat{\varrho}(t) = \int d[\xi] d[\eta]~
\langle \psi^-_{[\xi \eta]}(t)|\psi^+_{[\xi \eta]}(t) \rangle = 1, 
~~~~~~
\int d[\xi] d[\eta]~ = 1
\end{equation}
is satisfied by imposing
$\langle \psi^-_{[\xi \eta]}(t)|\psi^+_{[\xi \eta]}(t) \rangle = 1$.
This fixes the arbitrary functions $a(t)$ and $b(t)$.
Indeed, the requirement that the Ito differential 
\begin{eqnarray}
d ~\langle \psi^-_{[\xi \eta]}(t)|\psi^+_{[\xi \eta]}(t) \rangle &=&
\langle \psi^-_{[\xi \eta]}(t)| \hat{A} - a(t) + \hat{A}^{\dag} - a^*(t)
| \psi^+_{[\xi \eta]}(t) \rangle ~ dw_\xi(t)
\nonumber \\ &&+ 
2 i \langle \psi^-_{[\xi \eta]}(t)| \hat{B} - b(t)
| \psi^+_{[\xi \eta]}(t) \rangle ~ dw_\eta(t)
\end{eqnarray}
vanishes, implies
\begin{eqnarray}
a(t) &=& 
\langle \psi^-_{[\xi \eta]}(t)| \hat{A} |\psi^+_{[\xi \eta]}(t) \rangle 
\nonumber \\
&=& \sqrt{2m \gamma k_BT \over \hbar^2} ~
\langle \psi^-_{[\xi \eta]}(t)| \hat{q} |\psi^+_{[\xi \eta]}(t) \rangle 
+ i \sqrt{\gamma \over 8 m k_BT} ~
\langle \psi^-_{[\xi \eta]}(t)| \hat{p} |\psi^+_{[\xi \eta]}(t) \rangle 
\\
b(t) &=&
\langle \psi^-_{[\xi \eta]}(t)| \hat{B} |\psi^+_{[\xi \eta]}(t) \rangle
= \sqrt{\gamma \over 8 m k_BT} ~
\langle \psi^-_{[\xi \eta]}(t)| \hat{p} |\psi^+_{[\xi \eta]}(t) \rangle. 
\end{eqnarray}

The appearance of two different Green functions $G^\pm_{[\xi \eta]}$ 
in Eq. (\ref{RHODECT}) possibly introduces a violation of
positivity in $\hat{\varrho}(t)$.
This unphysical property is reflected in the anomalous definition
of expectation values of Hermitian operators, e.g., 
$\langle \psi^-_{[\xi \eta]}(t)| \hat{q} |\psi^+_{[\xi \eta]}(t) \rangle$,
which may be complex.
The problem is mathematically related to the presence of the $B$ terms
in Eq. (\ref{ZPREOPEN}). 
However, due to the high temperature condition $\hbar/k_BT \ll \tau$
we have $B \ll A$ (see Appendix A for details) and the last square 
in Eq. (\ref{ZPREOPEN}) can be neglected with respect to the last but one.
In this case, Eq. (\ref{CLMEQ}) becomes
\begin{eqnarray}
{d\over dt} \hat{\varrho}(t) &=&
-{i\over\hbar} \left[ \hat{H}(\hat{p},\hat{q},t),\hat{\varrho}(t) \right]
\nonumber \\ &&
- {m \gamma k_BT \over \hbar^2}
\left[ \hat{q}, \left[ \hat{q},\hat{\varrho}(t) \right] \right]
- {i \gamma \over 2 \hbar}
\left[ \hat{q}, \left\{ \hat{p},\hat{\varrho}(t) \right\} \right]
- {\gamma \over 16 mk_BT}
\left[ \hat{p}, \left[ \hat{p},\hat{\varrho}(t) \right] \right]
\nonumber \\ &=&
- {i \over \hbar} \left[ \hat H(\hat p,\hat q,t) +
{\gamma \over 4} (\hat{p}\hat{q}+\hat{q}\hat{p}), \hat{\varrho}(t) \right]
+ {1\over2} \left[ \hat{A} \hat{\varrho}(t), \hat{A}^{\dag} \right]
+ {1\over2} \left[ \hat{A}, \hat{\varrho}(t) \hat{A}^{\dag} \right].
\label{PCLMEQ}
\end{eqnarray}
This equation is of Lindblad class and provides a (completely)
positive evolution of $\hat{\varrho}(t)$ \cite{LINDBLAD}.
The reduced density matrix operator can be decomposed in terms 
of a single state $| \psi_{[\xi]}(t) \rangle$ associated to the 
Green function $G_{[\xi]}$ obtained by neglecting the $B$ terms in 
$| \psi^\pm_{[\xi \eta]}(t) \rangle$ and $G^\pm_{[\xi \eta]}$, respectively.
Note the disappearance of the $\eta$ noise.
Equation (\ref{STOCEQ}) becomes the norm-preserving 
stochastic Schr\"odinger equation 
\begin{eqnarray}
d | \psi_{[\xi]}(t) \rangle &=& 
- {i \over \hbar} \left[ \hat H(\hat p,\hat q,t) +
{\gamma \over 4} (\hat{p}\hat{q}+\hat{q}\hat{p}) \right]
| \psi_{[\xi]}(t) \rangle~ dt
\nonumber \\ &&
-{1\over2} \left[ \hat{A}^{\dag}\hat{A} + a(t)^*a(t) - 2a(t)^*\hat{A}\right]
| \psi_{[\xi]}(t) \rangle~ dt
+\left[ \hat{A} - a(t) \right] | \psi_{[\xi]}(t) \rangle~ dw_\xi(t)
\label{PSTOCEQ}
\end{eqnarray}
with $a(t)= \langle \psi_{[\xi]}(t)| \hat{A} |\psi_{[\xi]}(t) \rangle$.

Equation (\ref{PSTOCEQ}) describes the evolution of the measured system
for a realization of the stochastic processes 
$|\psi_{[\xi]}(t) \rangle$. 
This is a selective measurement process which is related the nonselective 
one by the relationship  
\begin{equation}
\hat{\rho}(t) = \overline{|\psi_{[\xi]}(t) \rangle \langle \psi_{[\xi]}(t)|}
= \int d[\xi]~|\psi_{[\xi]}(t) \rangle \langle \psi_{[\xi]}(t)|.
\label{PSIPSISUM}
\end{equation}
The quantum expectation values of the observables, e.g., $q(t)=
\langle \psi_{[\xi]}(t)| \hat{q} |\psi_{[\xi]}(t) \rangle $,
are stochastic processes with average value
\begin{equation} 
\overline{q(t)} = \int d[\xi]~
\langle \psi_{[\xi]}(t)| \hat{q} |\psi_{[\xi]}(t) \rangle 
\end{equation}
and variance
\begin{equation} 
\Delta q^2(t) = \int d[\xi]~
\langle \psi_{[\xi]}(t)| \left[ \hat{q} - \overline{q(t)} \right]^2
|\psi_{[\xi]}(t) \rangle.
\end{equation}
According to Eq. (\ref{PSIPSISUM}),
these average quantities can be also directly computed by considering 
the nonselective measurement process described by Eq. (\ref{PCLMEQ})
\begin{equation} 
\overline{q(t)} = \mbox{Tr} \left[ \hat{\varrho}(t) \hat{q} \right] 
= \int dpdq~W(p,q,t)~q
\label{QAVE}
\end{equation}
\begin{equation} 
\Delta q^2(t) = \mbox{Tr} \left\{ \hat{\varrho}(t)
\left[ \hat{q}-\overline{q(t)} \right]^2 \right\} =
\int dpdq~W(p,q,t)~\left[ q - \overline{q(t)} \right]^2.
\label{QVAR}
\end{equation}
The last expressions, formally identical to the classical Eqs. 
(\ref{CAVE}) and (\ref{CVAR}), have been obtained by introducing 
the Wigner function $W(p,q,t)$ through the relation 
$\varrho(q,q,t)=\int dp~W(p,q,t)$.
For $\gamma \to 0$, the effect of the measurement vanishes. 
In this case, Eq. (\ref{PSTOCEQ}) becomes the Schr\"odinger equation 
for the isolated system and the variance (\ref{QVAR}) reduces to 
the standard quantum mechanical expression. 

Finally, we show that the von Neumann collapse theory is recovered 
for $\gamma \to \infty$ and $T\to \infty$  with $T\gamma^{-1}$ constant.
In this limit, we must identify the shortest time scale $\tau$ of the
classical system with $\gamma^{-1}$ and the condition $T \gamma^{-1}$
constant allows the inequality $\hbar/k_BT \ll \Omega^{-1} \ll \tau$
to be always satisfied.
At time $t=t'+\tau$, Eqs. (\ref{PCLMEQ}) and (\ref{PSTOCEQ}) give
\begin{eqnarray}
\varrho(q_1,q_2,t'+\tau) &\simeq& 
\exp\left[-{1 \over 2}\kappa\tau (q_1-q_2)^2\right]~\varrho(q_1,q_2,t')
\label{VN1} \\ 
\psi_{\xi'}(q,t'+\tau) &\simeq&  
\exp \left[ -\kappa\tau \left( q-q'-{\xi'\over 2\sqrt{\kappa}} \right)^2 
\right] ~\exp \left( {\xi'^2 \tau \over 4} \right)~ \psi(q,t'),
\label{VN2}
\end{eqnarray}
where $\kappa=2m \gamma k_B T/\hbar^2$, $\xi'=\xi(t')$, and  $q'=q(t')$.
For $T\gamma^{-1}$ constant and $\tau = \gamma^{-1} \to 0$, 
Eqs. (\ref{VN1}) and (\ref{VN2}) provide an instantaneous diagonalization  
of the reduced density matrix and an instantaneous collapse of the 
wavefunction into the eigenfunction of $\hat{q}$
corresponding to the eigenvalue $q'+\xi'/2\sqrt{\kappa}$.
The normalization factor in Eq. (\ref{VN2}) is such that the quantum 
expectation values at time $t'+\tau$, when averaged over the noise 
realizations $\xi'$, coincide with the quantum expectation values at time 
$t'$.
For instance, we have
\begin{eqnarray} 
\overline{ 
\langle \psi_{\xi'}(t'+\tau)| \hat{q} |\psi_{\xi'}(t'+\tau) \rangle}
&=& \int d\xi' \sqrt{\tau \over 2\pi} \exp\left( -{\xi'^2 \tau \over 2}\right)
~ \int dq~q~\left| \psi_{\xi'}(q,t'+\tau) \right|^2
\nonumber \\ &=& 
\int dq ~q~ |\psi(q,t')|^2
\end{eqnarray}
which is the result expected on the basis of the von Neumann postulate:
in a selective measurement of position at time $t'$,
the probability that the state $|\psi(t')\rangle$ collapses into 
the eigenstate $|q\rangle$ is $|\langle q|\psi(t')\rangle|^2$.

\section{Measurement results}

In the previous Sections, we have seen how the evolution of a system, 
classical or quantum, is influenced by coupling its coordinate to those of 
infinitely many linear oscillators.
We called this process a measurement of position in agreement with the fact 
that in a proper limit the von Neumann collapse theory can be recovered from
the resulting equations.
Now, we specify how the properties of the measured system can be 
operatively read by the observer.

From the point of view of the observer the linear oscillators of the 
meter always have classical features so that their coordinates can be 
directly taken as pointers of the meter itself.
Consider, for example, a pointer whose value $R(t)$ is defined as  
\begin{equation}
R(t)= \int ds \sum_n \zeta_n(t-s) Q_n(s).
\label{R}
\end{equation}
During a single measurement, i.e., in a selective process, $R(t)$ is a
stochastic variable. 
We would like to choose the response functions $\zeta_n(t-s)$ in order that
the statistical properties of $R(t)$ over the ensemble of all possible 
measurements coincide or, at least, allow us to recover the nonselective 
properties of the measured system. 
For instance, we could ask that the average measurement result
\begin{equation}
\overline{R(t)}=\int ds \sum_n \zeta_n(t-s) \overline{Q_n(s)} 
\end{equation}
and its variance
\begin{equation}
\Delta R^2(t) = \int ds \sum_n \zeta_n(t-s) \left[ \overline{Q_n(s)^2} - 
\overline{Q_n(s)}^2 \right],
\end{equation}
correspond to the quantities (\ref{CAVE}) and (\ref{CVAR}) in the case
of a classical system or those (\ref{QAVE}) and (\ref{QVAR}) in the case
a quantum one.

In the classical case, Eq. (\ref{CQSOL}) provides an explicit expression
of the oscillator coordinates. 
By using the property $\Omega^{-1} \ll \tau$, we see that for 
$\omega_n \sim \Omega$ we can find 
a period $\lambda^{-1}$ much shorter than the fastest classical 
time and much longer than the inverse of the oscillator frequency 
so that the average of $Q_n(t)$ in this period coincides with $q(t)$.
With the choice 
$\zeta_n(t) = \lambda \exp(-\lambda t)~\delta_{\omega_n-\Omega}$
and using the definition (\ref{CAVEDEF}) we then have 
\begin{equation}
\overline{R(t)} = \overline{q(t)}
\label{RAVE}
\end{equation}
and
\begin{equation}
\Delta R^2(t) = \Delta q^2(t) + \ell^2.
\label{RVAR}
\end{equation}
The pointer variance is the sum of the variance $\Delta q^2(t)$ of the
measured system and the resolution of the measurement apparatus
\begin{equation}
\ell^2 = {k_BT \over M\Omega^2}. 
\label{RESOLUTION}
\end{equation}
The term $\ell^2$ represents a systematic error of the measurement and can,
in principle, be subtracted.

The above results can be derived in an alternative way.
Consider the general definition of the moments
\begin{equation}
\overline{Q_n(t)} = \int dpdqdPdQ~W_{tot}(p,q,P,Q,t)~Q_n 
\label{CQAVE}
\end{equation}
\begin{equation}
\overline{Q_n(t)^2} = \int dpdqdPdQ~W_{tot}(p,q,P,Q,t)~Q_n^2.
\label{CQVAR}
\end{equation}
Here, $W_{tot}(p,q,P,Q,t)$ is the probability density  
solution of the Liouville equation for the total system with initial 
conditions
\begin{equation}
W_{tot}(p,q,P,Q,t')= \delta(p-p') \delta(q-q')
{ e^{- H_m(P',Q'-q') / k_BT} \over \int dQ'dP'~e^{- H_m(P',Q'-q') / k_BT} }.
\end{equation}
The oscillators with frequency $\omega_n \sim \Omega$ approach the 
thermal equilibrium around the instantaneous value of the measured coordinate
on a time scale much shorter than a characteristic period $\lambda^{-1}$
with $\Omega^{-1} \ll \lambda^{-1} \ll \tau$.
For these oscillators the time average of the moments (\ref{CQAVE}) and 
(\ref{CQVAR}) over a period $\lambda^{-1}$ can be approximated by inserting 
the following adiabatic expression for the total probability density 
in the same Eqs (\ref{CQAVE}) and (\ref{CQVAR})
\begin{equation}
W_{tot}(p,q,P,Q,t) \simeq W(p,q,t) ~
{ e^{- H_m(P,Q-q(t)) / k_BT} \over \int dQ dP~e^{- H_m(P,Q-q(t)) / k_BT} }, 
\label{ADIAB}
\end{equation}
where $W(p,q,t)$ is the solution of Eq. (\ref{FOKKERPLANK}).
Equations (\ref{RAVE}) and (\ref{RVAR}) are then obtained by evaluating
the Gaussian integrals in (\ref{CQAVE}) and (\ref{CQVAR}).

The last approach formally applies also at quantum level by interpreting 
the $W$s as Wigner functions.
The adiabatic approximation (\ref{ADIAB}) becomes
\begin{equation}
W_{tot}(p,q,P,Q,t)\simeq W(p,q,t)~W_m(P,Q-q(t)),
\end{equation}
where $W(p,q,t)$ is the Wigner function associated to $\hat{\varrho}(t)$
and 
\begin{equation}
W_m(P,Q-q(t)) = \prod_{ n} 
{\tanh \left( {\hbar \omega_{n} \over 2 k_B T} \right) \over \pi \hbar}
\exp \left\{ - 
\tanh \left( {\hbar \omega_{n} \over 2 k_B T} \right) 
\left[ {P_n^2 \over \hbar M \omega_n} + {M \omega_n \over \hbar}
\left( Q_n-q(t) \right)^2 \right] \right\}
\end{equation}
is the Wigner function associated to the density matrix (\ref{RHOENV})
with $Q_{eq}=q(t)$.
Here, $q(t)$ is the quantum expectation value of $\hat{q}$ in the 
state $|\psi_{[\xi]}(t) \rangle$.
Due to the condition $k_BT \gg \hbar \Omega$, the above expression 
reduces to the classical distribution 
\begin{equation}
W_m(P,Q-q(t)) = \prod_{ n} { \omega_n \over 2 \pi k_BT}
\exp \left\{ - 
\left[ {P_n^2 \over 2 M} + {M \omega_n^2 \over 2}
\left( Q_n-q(t) \right)^2 \right]{1\over k_BT} \right\}
\end{equation}
so that Eqs. (\ref{RAVE}) and (\ref{RVAR}) still hold 
with the same resolution $\ell$ of the measurement apparatus.

\section{The $\protect\bbox{\hbar \to 0}$ limit: definition of coherent states}

The dynamics of a closed quantum system reduces to the classical dynamics in 
the $\hbar \to 0$ limit only if the system is prepared in appropriate states.
The coherent states defined as the ground state of the displaced harmonic 
oscillator 
\begin{equation}
|p'q' \rangle = 
e^{-{i\over\hbar} q' \hat{p}} 
e^{ {i\over\hbar} p' \hat{q}} | \phi_0 \rangle,
\label{CS}
\end{equation}
where $| \phi_0 \rangle$ is the ground state of the undisplaced oscillator,
are a well known example \cite{KLAUDER-SUDARSHAN}.
These states provide a convenient representation for studying the 
$\hbar \to 0$ limit regardless of the nature of the Hamiltonian 
which may not preserve their form \cite{YAFFE}. 

In the case of the measurement model discussed here, it is possible
to find states which are localized and stationary in the comoving frame
of the measured system and play the role of the ground state
$| \phi_0 \rangle$ in Eq. (\ref{CS}). 
A first example of these states was given in \cite{DIOSI88B,GRW} 
for a free particle evolving according to the dissipationless Eq. (\ref{E2}). 
A generalization valid in the case of a harmonic oscillator described by 
Eq. (\ref{PSTOCEQ}) has been recently provided in \cite{HALLIWELL95B}.  
Here, we derive the expression of the coherent states for a general linear 
system with constant proper frequency undergoing the selective measurement 
process of Eq. (\ref{PSTOCEQ}).
Then, we discuss the recovering of the classical limit in selective and 
nonselective measurement processes on arbitrary systems which are prepared 
in such states.

During a selective measurement, the quantum system is described by  
a state $|\psi_{[\xi]}(t) \rangle$ which evolves in a specified rest-frame
according to Eq. (\ref{PSTOCEQ}).
In analogy with Eq. (\ref{CS}), we look for solutions of 
Eq. (\ref{PSTOCEQ}) of the form 
\begin{equation}
|\psi_{[\xi]}(t) \rangle = 
e^{-{i\over\hbar} q(t) \hat{p}} 
~e^{ {i\over\hbar} p(t) \hat{q}} 
~ e^{ -{i\over\hbar} \varphi(t)} |\phi\rangle,
\label{TRANSFORM}
\end{equation}
where
$p(t) = \langle \psi_{[\xi]}(t)| \hat{p} |\psi_{[\xi]}(t) \rangle$
and $q(t) = \langle \psi_{[\xi]}(t)| \hat{q} |\psi_{[\xi]}(t) \rangle$.
The co-moving state $|\phi\rangle$ is assumed constant so that the 
solutions (\ref{TRANSFORM}) depend on the noise $\xi(t)$
only through the expectation values $p(t)$ and $q(t)$ and the action
$\varphi(t)$.
By inverting the transformation (\ref{TRANSFORM}) and imposing that 
the change of $|\phi\rangle$ in a time $dt$ vanishes, we get 
\begin{equation}
e^{{i\over\hbar} [\varphi(t) +d\varphi(t)]}
~e^{-{i\over\hbar} [p(t) +dp(t)] \hat{q}}
~e^{{i\over\hbar} [q(t)+dq(t)] \hat{p}} 
\left[ |\psi_{[\xi]}(t) \rangle + d|\psi_{[\xi]}(t) \rangle \right]
- |\phi \rangle =0.
\label{CHANGE}
\end{equation}
The differential $d|\psi_{[\xi]}(t) \rangle$ is given by Eq. (\ref{PSTOCEQ}).
The same Eq. (\ref{PSTOCEQ}) allows the evaluation of  
\begin{equation}
dp(t) = - \left[ \gamma p(t) +  
\langle \psi_{[\xi]}(t)| \partial_q\hat{V}(\hat{q},t) |\psi_{[\xi]}(t) 
\rangle \right] dt 
+ 2\sqrt{\kappa} \sigma_{pq}^2 dw_\xi(t)
\label{QLANGE1} 
\end{equation}
and 
\begin{equation}
dq(t) = {p(t) \over m} dt 
+ \left( 2\sqrt{\kappa}\sigma_{q}^2 - {\gamma\over 2\sqrt{\kappa}}
\right) dw_\xi(t),
\label{QLANGE2}
\end{equation}
where 
\begin{eqnarray}
\sigma_q^2 &=& 
\langle \psi_{[\xi]}(t)| \hat{q}^2 |\psi_{[\xi]}(t) \rangle
-\langle \psi_{[\xi]}(t)| \hat{q} |\psi_{[\xi]}(t) \rangle^2
=\langle \phi| \hat{q}^2 |\phi \rangle
\label{SIGMAQ} 
\end{eqnarray}
and 
\begin{eqnarray}
\sigma_{pq}^2 &=& {1\over2} 
\langle \psi_{[\xi]}(t)| \hat{p}\hat{q}+\hat{q}\hat{p} 
|\psi_{[\xi]}(t) \rangle
-\langle \psi_{[\xi]}(t)| \hat{p} |\psi_{[\xi]}(t) \rangle
\langle \psi_{[\xi]}(t)| \hat{q} |\psi_{[\xi]}(t) \rangle
= {1\over2} 
\langle \phi| \hat{p}\hat{q}+\hat{q}\hat{p} |\phi \rangle
\label{SIGMAPQ}
\end{eqnarray}
are the constant variances associated to the states (\ref{TRANSFORM})
and $\kappa = 2m\gamma k_BT /\hbar^2$.
The expectation value of the force operator which appears in Eq. 
(\ref{QLANGE1}) can be expressed in terms of the co-moving state 
$|\phi\rangle$ by a translation $\hat{q} \to \hat{q}+q(t)$ 
\begin{equation}
\langle \psi_{[\xi]}(t)| \partial_q\hat{V}(\hat{q},t) |\psi_{[\xi]}(t) 
\rangle =  
\langle \phi| \partial_q\hat{V}(\hat{q}+q(t),t) |\phi \rangle.
\end{equation}
Finally, we write the differential of the action $\varphi(t)$ 
in terms of two coefficients $\mu(t)$ and $\nu(t)$ to be determined later
\begin{equation}
d\varphi(t) = \mu(t) dt + \nu(t) dw_\xi(t).
\end{equation}
By expanding the exponentials and using the Ito rule, Eq. (\ref{CHANGE}) 
can be rewritten as
\begin{equation}
\left[ \hat{1} + \hat{F} dw_\xi(t) + \hat{G} dt \right] |\phi \rangle 
- | \phi \rangle = 0,
\end{equation}
which is equivalent to $\hat{F} |\phi \rangle =0$ and 
$\hat{G} |\phi \rangle =0$.
In general, the operators $\hat{F}$ and $\hat{G}$ will depend on time 
through the expectation values $p(t)$ and $q(t)$ and the action $\varphi(t)$
so that these equations cannot be satisfied with a constant 
$|\phi \rangle$.
However, we can try to make $\hat{F}$ and $\hat{G}$ time independent
with a proper choice of the coefficients $\mu(t)$ and $\nu(t)$.
In the case of $\hat{F}$, we have  
\begin{equation}
\hat{F} = \sqrt{\kappa} 
\left[\left( 1 - {2i\over\hbar} \sigma_{pq}^2 \right) \hat{q}
+ {2i\over\hbar} \sigma_q^2 \hat{p} \right]
+ {i\over \hbar} \left[ \nu(t) + p(t) \left( 2 \sqrt{\kappa} \sigma_q^2
- {\gamma \over 2 \sqrt{\kappa}} \right) \right] 
\end{equation}
and this becomes time independent with the choice
\begin{equation}
\nu(t) = - p(t) \left( 2 \sqrt{\kappa} \sigma_q^2
- {\gamma \over 2 \sqrt{\kappa}} \right).  
\label{NU}
\end{equation}
The corresponding equation $\hat{F} |\phi \rangle = 0$ 
has the unique normalized solution
\begin{equation}
\phi(q) = \langle q | \phi \rangle =
\left( 2 \pi \sigma_q^2 \right)^{-1/4}~ \exp \left( - {1 - {2i\over \hbar}
\sigma_{pq}^2 \over 4 \sigma_q^2}~ q^2 \right).
\label{PHIS}
\end{equation}
For $\nu(t)$ given by Eq. (\ref{NU}), the operator $\hat{G}$ is
\begin{eqnarray}
\hat{G} =  - {i \over \hbar} &\bigg[& {\hat{p}^2 \over 2m} 
+ \hat{V}(\hat{q}+q(t),t) 
- \langle \phi| \partial_q\hat{V}(\hat{q}+q(t),t) |\phi \rangle \hat{q}
- \mu(t) - {p(t)^2 \over 2m} + {\gamma \over 2}p(t)q(t) 
\nonumber \\ &&
+ {\gamma \over 2} (\hat{p}\hat{q}+\hat{q}\hat{p}) 
- 2\kappa \sigma_q^2 \sigma_{pq}^2 \bigg]
- \kappa \left(\hat{q}^2 - \sigma_q^2 \right) 
+ {1\over 2} \hat{F}^2.
\end{eqnarray}
This can be made time independent with a proper choice of 
$\mu(t)$ only for linear systems.
Assuming $V(q,t)=v_0(t)+v_1(t)q+\case{1}{2}m\omega_0^2q^2$ with $\omega_0$
constant, we have 
\begin{equation}
\hat{V}(\hat{q}+q(t),t) 
- \langle \phi| \partial_q\hat{V}(\hat{q}+q(t),t) |\phi \rangle \hat{q}
= V(q(t),t) + {1\over 2}m\omega_0^2\hat{q}^2 
\end{equation}
so that by choosing 
\begin{equation}
\mu(t) = \epsilon -{p(t)^2 \over 2m} + V(q(t),t) + {\gamma \over 2}p(t)q(t), 
\label{MU}
\end{equation}
the equation $\hat{G} |\phi \rangle=0$ becomes
\begin{equation}
\left[ {\hat{p}^2 \over 2m} + {1\over 2}m\omega_0^2\hat{q}^2 
+{\gamma \over 2} (\hat{p}\hat{q}+\hat{q}\hat{p}) 
- 2\kappa \sigma_q^2 \sigma_{pq}^2
- i\hbar \kappa \left(\hat{q}^2 - \sigma_q^2 \right) 
\right] | \phi \rangle = \epsilon | \phi \rangle.
\label{CMSS2}
\end{equation}
In the position representation and using $\phi(q)$ given by (\ref{PHIS}), 
Eq. (\ref{CMSS2}) is equivalent to the following two complex equations
\begin{eqnarray}
-{\hbar^2\over 2m}~\left({1 - {2i\over \hbar} \sigma_{pq}^2 
\over 2 \sigma_q^2} \right)^2 
+ {m \omega_0^2 \over 2}
+ {i\hbar\gamma\over 2}~{1 - {2i\over \hbar} \sigma_{pq}^2 \over  \sigma_q^2}
- i \hbar \kappa &=& 0
\label{C1} \\
{\hbar^2\over 2m} {1 - {2i\over \hbar} \sigma_{pq}^2 \over 2 \sigma_q^2}
- {i\hbar\gamma \over 2} -2\kappa \sigma_q^2 \sigma_{pq}^2
+ i \hbar \kappa \sigma_q^2 &=& \epsilon.
\label{C2}
\end{eqnarray}
Equation (\ref{C1}) gives two conditions for the determination 
of $\sigma_q^2$ and $\sigma_{pq}^2$.
The constant $\epsilon$ can be then evaluated from the real part of 
Eq. (\ref{C2}), the imaginary part being an identity.
The solutions are 
\begin{eqnarray}
\sigma_q^2 &=& \sqrt{\gamma^2 -\omega_0^2 + 
\sqrt{(\gamma^2-\omega_0^2)^2 + \left(2\hbar\kappa/m\right)^2}
\over 8 \kappa^2}, 
\label{SIGMAQSOL} \\
\sigma_{pq}^2 &=& \sqrt{m^2(\gamma^2-\omega_0^2) \sigma_q^4 + 
{\hbar^2\over 4}} - m\gamma \sigma_q^2, 
\label{SIGMAPQSOL}
\end{eqnarray}
and 
\begin{eqnarray}
\epsilon = {\hbar^2 \over 4m \sigma_q^2} - 2\kappa \sigma_q^2 \sigma_{pq}^2.
\end{eqnarray}
The variances $\sigma_q^2$ and $\sigma_{pq}^2$ are always real and positive
except for $k_BT/\hbar\omega_0 \ll 1$ which is, however, outside the range 
of validity of Eq. (\ref{PSTOCEQ}). 
According to the choices (\ref{NU}) and (\ref{MU}), we finally have 
\begin{equation}
d\varphi(t) = \left[ \epsilon + {p(t)^2 \over 2m} + V(q(t),t) + 
{\gamma \over 2}p(t)q(t) \right] dt - p(t)dq(t).
\label{DPHI}
\end{equation}
This allows for an interpretation of $\epsilon$ in terms of a zero-point
energy which adds to the classical renormalized Hamiltonian
$H(p(t),q(t),t)+\case{\gamma}{2}p(t)q(t)$.

In the $\gamma \to 0$ limit, we have $\sigma_q^2 = \hbar / 2m\omega_0$,
$\sigma_{pq}^2 =0$, and $\epsilon=\hbar \omega_0 /2$.
The stationary state of Eq. (\ref{PHIS}), becomes the ground state 
$|\phi_0 \rangle$ of an unmeasured harmonic oscillator with frequency 
$\omega_0$.
In analogy with Eq. (\ref{CS}), the coherent states in presence of a
continuous measurement are then defined as
\begin{equation}
|p'q' \rangle = 
e^{-{i\over\hbar} q' \hat{p}} 
e^{ {i\over\hbar} p' \hat{q}} | \phi \rangle, 
\label{MCS}
\end{equation}
which, in the position representation, becomes
\begin{equation}
\langle q | p'q' \rangle =
\left( 2 \pi \sigma_q^2 \right)^{-1/4}~ \exp \left[ - {1 - {2i\over \hbar}
\sigma_{pq}^2 \over 4 \sigma_q^2}~ (q-q')^2 +{i \over \hbar} p'(q-q') \right],
\label{IS1}
\end{equation}
with $\sigma_q^2$ and $\sigma_{pq}^2$ given by Eqs. (\ref{SIGMAQSOL})
and (\ref{SIGMAPQSOL}), respectively.
The states $|p'q'\rangle$ of Eq. (\ref{MCS}) have the same properties
of the usual coherent states \cite{KLAUDER-SUDARSHAN}.
In particular, they form an overcomplete basis with completeness relationship
\begin{equation}
\int {dp'dq' \over 2 \pi \hbar}~ | p'q' \rangle \langle p'q' | = \hat{1}
\end{equation}
and overlaps
\begin{equation}
\langle p'q'| p''q'' \rangle = \exp\left[ -C_{p'-p''q'-q''}
+ {i\over \hbar}~{p'+p'' \over 2} (q'-q'') \right],
\end{equation}
where
\begin{eqnarray}
C_{p'-p''q'-q''} &=& 
{\sigma_q^2 \over 2\hbar^2 }
\left[ (p'-p'') - {\sigma_{pq}^2 \over \sigma_q^2} (q'-q'') \right]^2
+{1\over 8 \sigma_q^2} (q'-q'')^2.
\label{CPQPQ}
\end{eqnarray}

Suppose that a quantum system, not necessarily a linear one, 
is prepared at time $t$ in the coherent state $| p(t) q(t) \rangle$.
To the leading order in $\hbar$, we have
\begin{equation}
\sigma_q^2 = \sqrt{\hbar^3 \over 8 m^2 \gamma k_BT}
\label{SQLO}
\end{equation}
and 
\begin{equation}
\sigma_{pq}^2 = {\hbar \over 2}.
\label{SPQLO}
\end{equation}
We also have 
\begin{eqnarray} 
\sigma_p^2 &=& \langle p(t) q(t) |\hat{p}^2 |p(t)q(t)\rangle
- \langle p(t) q(t) |\hat{p} |p(t)q(t)\rangle^2 =
\langle \phi| \hat{p}^2 |\phi \rangle 
\nonumber \\
&=& { {\hbar^2 \over 4} + \sigma_{pq}^4 \over \sigma_q^2}
= {\hbar^2 \over 2 \sigma_q^2} = \sqrt{2 m^2 \hbar \gamma k_BT}. 
\end{eqnarray} 
Note that these expressions are independent of $\omega_0$ \cite{NOTA2}.
Since $\sigma_p^2$, $\sigma_q^2$, and $\sigma_{pq}^2$ vanish for 
$\hbar \to 0$, in this limit the expectation values 
$p(t)=\langle p(t)q(t)| \hat{p} | p(t)q(t) \rangle$
and $q(t)=\langle p(t)q(t)| \hat{q} | p(t)q(t) \rangle$ 
can be interpreted as classical phase space coordinates.
In a selective measurement, their change
is given by Eqs. (\ref{QLANGE1}) and (\ref{QLANGE2}) 
\begin{eqnarray}
dp(t) &=& - \left[ \gamma p(t) + 
\langle p(t)q(t)| \partial_q\hat{V}(\hat{q},t) | p(t)q(t) \rangle \right] dt 
+ \sqrt{2m \gamma k_BT} dw_\xi(t)
\\
dq(t) &=& {p(t) \over m} dt 
+ \left( \sqrt{\hbar\over m} - \sqrt{\hbar^2 \gamma  \over 8mk_BT}
\right) dw_\xi(t).
\end{eqnarray}
For $\hbar \to 0$, the expectation value
$\langle p(t)q(t)| \partial_q\hat{V}(\hat{q},t) | p(t)q(t) \rangle$
can be replaced with $\partial_q V(q(t),t)$ and we recover the classical 
Langevin equations (\ref{LANGE1}) and (\ref{LANGE2}).

In the case of a nonselective measurement, the $\hbar \to 0$ limit is
properly discussed in terms of the Wigner function related to
the reduced density matrix operator through the transformation
\begin{equation}
W(p,q,t)= {1 \over 2 \pi \hbar} \int dz~ \exp \left({i\over\hbar} pz \right)
\langle q - {z\over 2}| \hat{\varrho}(t) |q+ {z\over 2} \rangle.
\end{equation}
Suppose that at time $t$ the measured system is described by the density 
matrix obtained by averaging the state $| p(t) q(t) \rangle$ over all noise 
realizations, i.e., all possible values of $p(t)$ and $q(t)$ specified
by a certain distribution function such that $\mbox{Tr} \hat{\varrho}(t)=1$,
\begin{equation}
\hat{\rho}(t) = \overline{ | p(t) q(t) \rangle \langle p(t) q(t) |}
= \int dp'dq'~\overline{\delta(p'-p(t))\delta(q'-q(t))} 
~| p'q' \rangle \langle p'q' |.
\end{equation}
The corresponding Wigner function is
\begin{equation}
W(p,q,t)= \int dp'dq'~\overline{\delta(p'-p(t))\delta(q'-q(t))} 
~W_{p'q'}(p,q),
\label{LAST}
\end{equation}
where
\begin{eqnarray}
W_{p'q'}(p,q) &=&
{1 \over 2 \pi \hbar} \int dz~ \exp \left({i\over\hbar} pz \right)
\langle q - {z\over 2}| p'q' \rangle \langle p'q' 
|q+ {z\over 2} \rangle
\nonumber \\ &=& 
{1\over \pi \hbar} \exp \left\{ -{2\sigma_q^2 \over \hbar^2}
\left[ (p-p') - {\sigma_{pq}^2 \over \sigma_q^2} (q-q') \right]^2
- {1\over 2 \sigma_q^2} (q-q')^2 \right\}.
\label{WPQPQ}
\end{eqnarray}
Since
\begin{equation} 
\lim_{\hbar \to 0}W_{p'q'}(p,q) = \delta(p-p') \delta(q-q'),
\end{equation}
in the $\hbar \to 0$ limit $W(p,q,t)$ reduces to the classical probability 
density 
$\overline{\delta(p-p(t))\delta(q-q(t))}$ obtained by averaging 
the sharp density $\delta(p-p(t))\delta(q-q(t))$ over all acceptable
phase space points $p(t)$, $q(t)$.
Finally, the equation of motion of the Wigner function, obtained
from Eq. (\ref{PCLMEQ}) with standard manipulations \cite{DV}, is 
\begin{eqnarray}
\partial_t W(p,q,t) = \biggl[ 
&-&{p \over m} \partial_q 
+\sum_{n=0}^{\infty} \left({\hbar \over 2i} \right)^{2n}
{1 \over (2n+1)!}\partial_q^{2n+1} V(q,t) \partial_p^{2n+1} 
\nonumber \\ &+&
\partial_p \left( \gamma p + m \gamma k_BT \partial_p \right) 
+ {\hbar^2 \gamma \over 16mk_BT} \partial_q^2 \biggr] W(p,q,t),
\label{PCLWEQ}
\end{eqnarray}
so that, in the $\hbar \to 0$ limit, the change of $W(p,q,t)$ coincides 
with that prescribed by the classical Fokker-Plank Eq. (\ref{FOKKERPLANK}).

\section{Measurements on macroscopic systems}

One of the principal drawbacks of the von Neumann measurement theory
is the impossibility of predicting a quantum-to-classical transition 
in the macroscopic limit, unless the state
$|\psi(t')\rangle$ of the system at the beginning of the measurement
is one of the coherent states (\ref{CS}).
On the other hand, 
when the size of the system is sufficiently large, 
i.e., in the formal $\hbar \to 0$ limit, we must always recover the result 
of a classical measurement. 
It is now clearly established that the entanglement of the measured system 
with the infinitely many degrees of freedom of the measurement apparatus
can provide superselection rules which avoid paradoxical quantum features
at macroscopic level \cite{ZUREK,CINI}.

Concerning the measurement model discussed here, the existence of a
superselection rule of this kind can be demonstrated in a general way 
in the case of linear systems. 
Halliwell and Zoupas have shown, in a statistical sense first
\cite{HALLIWELL95B} and with a more direct approach but in the 
free particle case and dissipationless limit later \cite{HALLIWELL96}, 
that the solutions of Eq. (\ref{PSTOCEQ}) converge to a coherent state 
characterized by time-dependent parameters $p(t)$ and $q(t)$ which are the 
expectation values of $\hat{p}$ and $\hat{q}$ in the state itself.
Here, we generalize the result of \cite{HALLIWELL96} by showing that 
the solutions of Eq. (\ref{PSTOCEQ}) with potential
\begin{equation} 
V=v_0(t)+v_1(t)q+\case{1}{2}m\omega_0^2 q^2
\label{QUADPOT}
\end{equation} 
in the long time limit become of the form  
\begin{equation}
|\psi_{[\xi]}(t)\rangle = \exp\left[-{i \over \hbar} \varphi(t) \right]
~|p(t)q(t)\rangle
\end{equation}
where 
$p(t)=\langle \psi_{[\xi]}(t)|\hat{p}| \psi_{[\xi]}(t) \rangle$, 
$q(t)=\langle \psi_{[\xi]}(t)|\hat{q}| \psi_{[\xi]}(t) \rangle$ and  
$\varphi(t)$ and $|p(t)q(t)\rangle$ are given by 
Eqs. (\ref{DPHI}) and (\ref{MCS}), respectively. 

The Green function corresponding to the nonlinear Eq. (\ref{PSTOCEQ}) 
\begin{eqnarray}
G_{[\xi]}(q,t;q',t') &=&
\int d[p] d[q]_{q',t'}^{q,t} 
\exp \bigg\{ \frac{i}{\hbar} \int_{t'}^t ds~ \bigg[
p\dot{q}-{p^2 \over 2m} -V - \gamma pq +\gamma p\langle\hat{q}\rangle
-{\gamma \over 2} \langle\hat{p}\rangle\langle\hat{q}\rangle
\nonumber \\&&
+i \hbar \kappa (q-\langle\hat{q}\rangle)^2
-i \hbar \sqrt{\kappa} (q-\langle\hat{q}\rangle) \xi
+{\gamma \over 2 \sqrt{\kappa}} (p-\langle\hat{p}\rangle) \xi 
\bigg]\bigg\}
\label{QUADGREEN}
\end{eqnarray}
functionally depends on the state $|\psi_{[\xi]}(t) \rangle$ through the
expectation values 
$\langle \hat{p}\rangle = 
\langle \psi_{[\xi]}(t)|\hat{p}| \psi_{[\xi]}(t) \rangle$ and 
$\langle \hat{q}\rangle = 
\langle \psi_{[\xi]}(t)|\hat{q}| \psi_{[\xi]}(t) \rangle$.
If we suppose, for the moment, that these functions and the noise $\xi$ 
are given, for $V$ of the form (\ref{QUADPOT}) the Green function 
(\ref{QUADGREEN}) is that of a linear system with Lagrangian 
\begin{equation}
L(q,\dot{q},t)={1\over 2} m \dot{q}^2 - {1\over 2} m \omega^2 q^2
- m \gamma q \dot{q} + f(t) q + g(t) \dot{q} + h(t),
\label{LAGRA}
\end{equation}
where
\begin{equation}
\omega^2 = \omega_0^2 - \gamma^2 -{2i \hbar \kappa \over m}
\label{OMELAGRA}
\end{equation}
and $f(t)$, $g(t)$, and $h(t)$ are given in terms of 
$\langle \hat{p}\rangle$, $\langle \hat{q}\rangle$, $\xi$, $v_0$, and $v_1$.
By performing the Gaussian functional integrals in Eq. (\ref{QUADGREEN}), 
we get 
\begin{eqnarray}
G_{[\xi]}(q,t;q',t') = n(t,t') \exp \bigg\{  {i\over \hbar} &\bigg[&
g(t)q-g(t')q' - {1\over 2} m\gamma q^2 + {1\over 2} m\gamma q'^2
+ S_{\text{cl}}(q,t;q',t') \bigg] \bigg\}, 
\label{GGG}
\end{eqnarray}
where $n(t,t')$ does not depend on the spatial variables and
\begin{eqnarray}
S_{\text{cl}}(q,t;q',t') &=& {1 \over \sin[\omega(t-t')]} \Bigg\{
{m \omega \over 2} \cos[\omega(t-t')] \left(q^2+q'^2\right) 
- m \omega qq'
\nonumber \\ &+& 
q \int_{t'}^t ds~ [f(s)-\dot{g}(s)] \sin[\omega(s-t')]
+q' \int_{t'}^t ds~ [f(s)-\dot{g}(s)] \sin[\omega(t-s)]
\nonumber \\ &-&
{1 \over m \omega } 
\int_{t'}^t ds~ [f(s)-\dot{g}(s)] \sin[\omega(t-s)]
\int_{t'}^s du~ [f(u)-\dot{g}(u)] \sin[\omega(u-t)] \Bigg\}
\label{SCL}
\end{eqnarray}
is the classical action of a driven harmonic oscillator of mass $m$, 
frequency $\omega$ and external force $f-\dot{g}$, evaluated 
with boundary conditions $q(t')=q'$ and $q(t)=q$ \cite{FEYNMAN}.
The frequency $\omega$ is complex with real and imaginary parts given by
\begin{equation}
\text{Re}(\omega) = \pm {\hbar \over 2 m}
\sqrt{8 \kappa^2 \over \gamma^2-\omega_0^2 + \sqrt{(\gamma^2-\omega_0^2)^2
+(2\hbar\kappa/m)^2}} = \pm {\hbar \over 2 m \sigma_q^2}
\label{REALOMEGA}
\end{equation}
\begin{equation}
\text{Im}(\omega) = \mp
\sqrt{\gamma^2-\omega_0^2 + \left({\hbar \over 2 m \sigma_q^2} \right)^2} 
\label{IMAGOMEGA}.
\end{equation}
For $(t-t') |\text{Im}(\omega)| \gg 1$, the coefficient of the $qq'$ 
term in the action (\ref{SCL}) vanishes while the coefficient of 
$(q^2+q'^2)$ becomes $\pm i m\omega/2$.
In the long time limit, therefore, the propagator (\ref{GGG}) looses 
memory of the 
initial conditions and the solutions of Eq. (\ref{PSTOCEQ}) can be written as
\begin{eqnarray}
\psi_{[\xi]}(q,t) &=& \int dq'~G_{[\xi]}(q,t;q',t')~\psi(q',t')
\nonumber \\ &=& \exp \left[ 
- {m\over 2\hbar} \left( \pm \omega + i \gamma  \right) q^2 
+\alpha(t) q + \beta(t) \right]
\nonumber \\ &=&
\exp \left[ - {1 - {2i\over \hbar} \sigma_{pq}^2 \over 4 \sigma_q^2}~q^2 
+\alpha(t) q + \beta(t) \right],
\label{ASSTAT}
\end{eqnarray}
where we used 
\begin{equation}
- {\hbar \text{Im}(\omega) \over 2 \text{Re}(\omega) }
\mp {\hbar\gamma \over 2\text{Re}(\omega)} = 
\sqrt{m^2(\gamma^2-\omega_0^2) \sigma_q^4 + 
{\hbar^2\over 4}} - m\gamma \sigma_q^2 = \sigma_{pq}^2. 
\end{equation}
The complex functions $\alpha(t)$ and $\beta(t)$ are to be determined.
First of all, the normalization of the wavefunction (\ref{ASSTAT}) 
implies that 
\begin{equation}
1 = (2\pi\sigma_q^2)^{1/2} \exp\left\{ 2\text{Re}[\beta(t)] + 
2\sigma_q^2 \text{Re}[\alpha(t)]^2 \right\}.
\end{equation}  
Then, we can impose two self-consistency conditions involving 
the expectation values of $\hat{p}$ and $\hat{q}$ in the state (\ref{ASSTAT}) 
\begin{equation}
p(t)= \langle \psi_{[\xi]}(t)|\hat{p}| \psi_{[\xi]}(t) \rangle
= \hbar\text{Im}[\alpha(t)] + 2\sigma_{pq}^2\text{Re}[\alpha(t)]
\end{equation}  
and
\begin{equation}
q(t)= \langle \psi_{[\xi]}(t)|\hat{q}| \psi_{[\xi]}(t) \rangle
= 2 \sigma_q^2\text{Re}[\alpha(t)].
\end{equation}  
By using the expressions so obtained for $\text{Re}[\alpha(t)]$, 
$\text{Im}[\alpha(t)]$, and $\text{Re}[\beta(t)]$, the wavefunction 
(\ref{ASSTAT}) can be rewritten as 
\begin{eqnarray}
\psi_{[\xi]}(q,t) &=& (2\pi\sigma_q^2)^{-1/4}
\exp \left\{ - {1 - {2i\over \hbar} \sigma_{pq}^2 \over 4 \sigma_q^2}~
[q-q(t)]^2 + {i\over \hbar} p(t)[q-q(t)] -{i\over\hbar}\varphi(t) \right\},
\label{LTPSI}
\end{eqnarray}
where 
\begin{equation}
\varphi(t)= -\hbar \text{Im}[\beta(t)] -p(t)q(t) + 
{\sigma_{pq}^2 \over 2\sigma_q^2}q(t)^2.
\end{equation}
The action $\varphi(t)$ evolves according to an equation obtained
by imposing that the change of $\psi_{[\xi]}(q,t)$ in a time $dt$
is given by Eq. (\ref{PSTOCEQ}).
Since the wavefunction (\ref{LTPSI}) is of the form (\ref{TRANSFORM}), 
the differential $d\varphi(t)$ is given by Eq. (\ref{DPHI}).
This completes the convergence proof.

To the leading order in $\hbar$, the characteristic time  which determines 
the convergence of $\psi_{[\xi]}(q,t)$ to the wavefunction (\ref{LTPSI})
is given by 
\begin{equation}
{1 \over |\text{Im}(\omega)|} = \sqrt{\hbar \over \gamma k_BT}.
\end{equation}
The convergence becomes infinitely fast for $\hbar \to 0$.
On the base of this result and of the properties of the coherent states 
discussed in the previous Section, we can conclude that during a measurement,
selective or nonselective, the $\hbar \to 0$ limit does exist 
at any time $t>t'$ even if it does not exist at $t=t'$. 
As an example of this behavior, in Appendix B we explicitly evaluate  
the $\hbar \to 0$ limit in the case 
of nonselective measurements on a free particle cat state.
The discontinuity at $t=t'$ is, of course, an artifact of the instantaneous 
correlation assumed through Eq. (\ref{RHOENV}) between the measured system 
and the measurement apparatus and it would disappear in a more physical 
approach in which such correlation is established in a finite time.

In the case of nonlinear systems, terms higher than quadratic appear 
in the potential of the Lagrangian (\ref{LAGRA}) so that the convergence
proof given for linear systems does not apply.
However, due to the linearity of the interaction with the infinitely 
many oscillators of the measurement apparatus, the leading $\hbar$ term of 
this potential, i.e.,  
\begin{equation}
-i\hbar \kappa q^2 = -i {2 m \gamma k_BT \over \hbar} q^2,
\end{equation}
is a quadratic one with complex frequency $\sqrt{-4i \gamma k_BT/\hbar}$.
As a consequence, in the $\hbar \to 0$ limit the state of the system becomes
of the form (\ref{LTPSI}) with $\sigma_q$ and $\sigma_{pq}$ given by 
Eqs. (\ref{SQLO}) and (\ref{SPQLO}).
The recovering of the classical behavior in the macroscopic limit is, 
therefore, obtained independently of the nature of the measured system.
Numerical examples of this result can be found in \cite{SPILLER,SAP,BRUN}
and experimental evidence has been recently reported in \cite{HAROCHE}.

After the completion of this paper, we became aware of a preprint by
Strunz and Percival \cite{STRUNZPERCIVAL} in which the authors discuss 
the semiclassical behavior of open quantum systems described by a general 
Lindblad master equation.

\acknowledgments
The authors are very grateful to Nicoletta Cancrini, Filippo Cesi, and Lajos
Di\'osi for useful discussions and to Walter Strunz for sending them the 
preprint \cite{STRUNZPERCIVAL}.
Partial support of INFM Sezione di Roma 1 is acknowledged.
M.P. was supported by the Laboratorio Forum, INFM Sezione di Firenze.

\appendix
\section{Positivity and Markovian evolution of $\protect\bbox{\hat{\varrho}}$}

Violations of the positivity of $\hat{\varrho}(t)$ may arise due to 
inappropriate approximations of the exact influence functional (\ref{GFI}).
Examples of exact solution of $\hat{\varrho}(t)$ with no positivity violations 
have been given in \cite{HUPAZZHANG} in the case of a harmonic oscillator. 

In the framework of a theory of measurement processes, 
the requirement that the meter has classical 
behavior with respect to the observer imposes $\hbar \Omega \ll k_BT$, 
if the frequency distribution (\ref{DENSITY}) is assumed.
In this case, the fluctuation kernel
\begin{eqnarray}
\Lambda(s-u) &=& \frac{m \gamma }{\pi \hbar} \int_0^{\Omega} d \omega~ 
\omega \coth \left( \frac{\hbar \omega}{2 k_B T} \right) 
\cos[\omega (s-u)]
\nonumber \\
&=& {k_BT \over \hbar^2} \Gamma(s-u)
- {1 \over 12 k_BT} \ddot{\Gamma}(s-u) + \ldots
\label{LAMBDA}
\end{eqnarray}
has a high temperature expansion whose leading term $\Gamma k_BT/\hbar^2$
is proportional to the dissipation kernel
\begin{equation}
\Gamma(t-s) \simeq 2 m \gamma \delta(t-s)
\end{equation}
which is Markovian for $\Omega^{-1} \ll \tau$, $\tau$ being the fastest
time scale of the classical motion.
When these approximations are made in Eq. (\ref{GFI}), so that 
Eq. (\ref{CLIF}) and the corresponding master equation (\ref{CLMEQ}) 
are obtained, violations of positivity of $\hat{\varrho}(t)$ may occur
as in the example pointed out in \cite{AMBEGAOKAR}.
However, this happens on a time scale shorter than
$\hbar/k_BT$, i.e., outside the range of validity 
$\tau \gg \Omega^{-1} \gg  \hbar/k_BT$ of Eq. (\ref{CLMEQ}) 
\cite{HUPAZZHANG}.
In this range, the substantial positivity of $\hat{\varrho}(t)$ can be 
made apparent by selecting appropriate dominant terms. 

Equation (\ref{CLIF}) shows that the dissipation term is negligible  
in comparison to the fluctuation one when
\begin{eqnarray}
{m \gamma k_BT q_\Delta^2 \over \hbar^2} \gg 
{\gamma q_\Delta p_\Sigma \over 2 \hbar},
\label{F>>D}
\end{eqnarray}
where $q_\Delta = q_1 - q_2$ and $p_\Sigma = p_1 + p_2$.
The functions $q_\Delta$ and $p_\Sigma$ may assume any value according
to the functional measure (\ref{DPDQ}).
However, close to the dominant classical path we have 
$p_\Sigma \lesssim 2 m q_\Delta/\tau$ and the condition (\ref{F>>D}) 
can be restated as $k_BT/\hbar \gg \tau^{-1}$.
Therefore, in the working range $\tau \gg \Omega^{-1} \gg  \hbar/k_BT$
dissipation can be neglected with respect to fluctuation 
and Eq. (\ref{CLIF}) becomes
\begin{eqnarray}
Z[q_1,q_2] =   
\int_{t'}^{t} ds ~ {m \gamma k_BT \over \hbar^2}
\left[ q_1(s) - q_2(s) \right]^2.
\end{eqnarray}
Correspondingly, the nonselective measurement processes are described
by the master equation
\begin{eqnarray}
{d\over dt} \hat{\varrho}(t) = 
-{i\over\hbar} \left[ \hat{H}(\hat{p},\hat{q},t),\hat{\varrho}(t) \right]
- {m \gamma k_BT \over \hbar^2}
\left[ \hat{q}, \left[ \hat{q},\hat{\varrho}(t) \right] \right].
\label{E1}
\end{eqnarray}
This is a Markovian evolution of Lindblad class  and therefore
(completely) positive \cite{LINDBLAD}.
The associated selective processes are described in 
terms of a single state satisfying the stochastic Schr\"odinger equation
\begin{eqnarray}
d | \psi_{[\xi]}(t) \rangle &=& 
- {i \over \hbar} \hat H(\hat p,\hat q,t)
| \psi_{[\xi]}(t) \rangle~ dt
\nonumber \\ &&
-{m \gamma k_B T \over \hbar^2} \left[ \hat{q} -q(t) \right]^2
| \psi_{[\xi]}(t) \rangle~ dt + \sqrt{2 m \gamma k_B T \over \hbar^2}
\left[ \hat{q} - q(t) \right] | \psi_{[\xi]}(t) \rangle~ dw_\xi(t),
\label{E2}
\end{eqnarray}
with $q(t) = \langle \psi_{[\xi]}(t) |\hat{q}| \psi_{[\xi]}(t) \rangle$.
The general results of \cite{PREONTA} are recovered
by setting $2m\gamma k_BT /\hbar^2 = \kappa$.

At classical level, the Fokker-Plank equation (\ref{FOKKERPLANK}) and 
the Langevin equations (\ref{LANGE1}) and (\ref{LANGE2}) are consistent 
with the fluctuation-dissipation theorem \cite{KUBO}. 
Equations (\ref{E1}) and (\ref{E2}), in which dissipation is neglected, 
 are therefore not appropriate for recovering the classical limit. 
New quantum equations are to be introduced which include  dissipation and,  
at the same time, guarantee the positivity of $\hat{\varrho}(t)$.
As shown in Section III, this is accomplished by rewriting
equation (\ref{CLIF}) in the equivalent form (\ref{ZPREOPEN}) 
and neglecting the $B$ terms with respect to the $A$ ones on the base
of the working condition $\hbar/k_BT \ll \tau$.
Dissipation is still contained in the remaining influence functional 
which gives rise to the master equation (\ref{PCLMEQ}) of Lindblad class 
and to the corresponding stochastic Schr\"odinger equation (\ref{PSTOCEQ}).
These equations provide the correct classical limit as shown in Section V.

We conclude with some remarks about the possibility
pointed out in \cite{DIOSI93A,DIOSI93B} of obtaining 
a master equation of Lindblad class by
taking into account the next to leading term in Eq. (\ref{LAMBDA}).
In this case, Eq. (\ref{CLIF}) would become
\begin{eqnarray}
Z[p_1,q_1,p_2,q_2] &=&   
\int_{t'}^{t} ds ~ \bigg\{ {m \gamma k_BT \over \hbar^2}
\left[ q_1(s) - q_2(s) \right]^2
+ {i \gamma \over 2 \hbar}  \left[ q_1(s) - q_2(s) \right]  
\left[ p_1(s) + p_2(s) \right] \bigg\} 
\nonumber \\ && 
- \int_{t'}^{t} ds \int_{t'}^s du~ {1 \over 12 k_BT}
\left[ q_1(s) - q_2(s) \right] \ddot{\Gamma}(s-u)
\left[ q_1(u) - q_2(u) \right]  + \ldots.
\end{eqnarray}
The new term can be more easily analyzed after integration by parts.
By setting $q_\Delta = q_1 - q_2$, we have
\begin{eqnarray}
&&\int_{t'}^{t} ds 
\int_{t'}^s du~ q_\Delta(s) \ddot{\Gamma}(s-u) q_\Delta(u)  
\nonumber \\ &&= 
{1\over 2} \int_{t'}^{t} ds~ \int_{t'}^t du~
 \ddot{q}_\Delta(s)  \Gamma(s-u) q_\Delta(u)
- {1\over 2} \Gamma(0) \left[ q_\Delta(t)^2  + q_\Delta(t')^2 \right]
+ \Gamma(t-t') q_\Delta(t) q_\Delta(t')
\nonumber \\ && 
+ {1\over 2} \int_{t'}^{t} du~ \Gamma(t-u) 
\left[ q_\Delta(t) \dot{q}_\Delta(u) - q_\Delta(u) \dot{q}_\Delta(t) \right] 
+ {1\over 2} \int_{t'}^{t} du~ \Gamma(t'-u) 
\left[ q_\Delta(u) \dot{q}_\Delta(t') - q_\Delta(t') \dot{q}_\Delta(u) \right].
\label{PARTINT}
\end{eqnarray}
For $t-t' \gg \Omega^{-1}$, the last three terms can be neglected and the 
first one approximated with a single integral
\begin{eqnarray}
{1\over 2} \int_{t'}^{t} ds~ q_\Delta(s) \ddot{q}_\Delta(s) =
{1\over 2} q_\Delta(t) \dot{q}_\Delta(t) -
{1\over 2} q_\Delta(t') \dot{q}_\Delta(t') -
{1\over 2} \int_{t'}^{t} ds~ \dot{q}_\Delta(s)^2. 
\end{eqnarray}
By using the identity 
\begin{eqnarray}
{1\over 2} \left[ q_\Delta(t)^2  + q_\Delta(t')^2 \right] 
=  q_\Delta(t')^2 + 
{1\over 2} \int_{t'}^{t} ds~ q_\Delta(s) \dot{q}_\Delta(s), 
\end{eqnarray}
Eq. (\ref{PARTINT}) can be rewritten as \cite{NOTAGAMMA}
\begin{eqnarray}
&&\int_{t'}^{t} ds~ 
\int_{t'}^s du~ q_\Delta(s) \ddot{\Gamma}(s-u) q_\Delta(u)  
\nonumber \\ &&= 
2 m \gamma \left\{ 
{1\over 2} \left[q_\Delta(t) \dot{q}_\Delta(t) - 
q_\Delta(t') \dot{q}_\Delta(t') \right] 
- {1\over 2} \int_{t'}^{t} ds~ \dot{q}_\Delta(s)^2 
- {\Omega \over \pi} q_\Delta(t')^2
- {\Omega \over \pi} \int_{t'}^{t} ds~ q_\Delta(s) \dot{q}_\Delta(s) \right\}.
\label{LTE}
\end{eqnarray} 
In Refs. \cite{DIOSI93A,DIOSI93B} the first term of Eq. (\ref{LTE}) is 
neglected by observing that it is much smaller than the third one.
In this case the reduced density matrix operator $\hat{\varrho}(t)$ would
undergo a transient change 
\begin{equation}
\varrho(q_1,q_2,0) \to 
\exp \left\{ - {2 m \gamma \Omega \over 12 \pi k_BT} 
\left[ q_1(t')-q_2(t') \right]^2 \right\} \varrho(q_1,q_2,0)
\label{TRANSIENT}
\end{equation}
followed by a Lindblad evolution described by an equation which reduces 
to Eq. (\ref{PCLMEQ}) in the $\hbar \to 0$ and $T \to \infty$ limits.
The validity of these findings is, however, questionable.
The surface terms neglected in Eq. (\ref{LTE}) are of the same order of 
the integral of $\case{1}{2} \dot{q}_\Delta^2$ which is, conversely, 
maintained.                 

The existence of a transient change in the evolution of the reduced
density matrix deserves further comments. 
If the system and the measurement apparatus are initially non-interacting,
a change of $\hat{\varrho}(t)$ at the switching-on of the interaction 
is plausible. 
However, as we explained in Section II, this transient cannot be described in 
the framework of a model, such as the bath of harmonic oscillators, 
in which the system and the measurement apparatus are always in interaction.
We must limit our considerations to a non-transient evolution and,
correspondingly, assume that the system and the meter
are correlated from the beginning.
 
The drawbacks of Refs. \cite{DIOSI93A,DIOSI93B} have been recently 
underlined also in \cite{TAMESHTIT}. 
Similarly to Ref. \cite{PATRIARCA} and the present work, 
the authors of \cite{TAMESHTIT} assume an initial correlation between 
the system and the environment.

\section{Nonselective measurements on a free particle cat state}

Let us consider a free quantum particle that, at the beginning of the
measurement, is in the superposition (cat) state
\begin{equation}
|\psi(t') \rangle = N \Big[ |p_1q_1\rangle + |p_2q_2\rangle \Big],
\label{CAT}
\end{equation}  
where $|p_iq_i\rangle$, $i=1,2$, are two coherent states (\ref{MCS})
with $\omega_0=0$ and the normalization factor is
\begin{equation}
N = {1\over\sqrt{2 \left[ 1 + \exp\left(-C_{p_1-p_2q_1-q_2}\right)\right] }}.
\end{equation}
The initial state (\ref{CAT}) has no classical counterpart.
The $\hbar \to 0$ limit of the corresponding Wigner function
\begin{eqnarray}
W(p,q,t') = N^2 &\Bigg\{& W_{p_1q_1}(p,q)+W_{p_2q_2}(p,q) 
+W_{{p_1+p_2\over 2}{q_1+q_2 \over 2}}(p,q)~
\nonumber \\&&\times
2 \cos \Bigg[ {p \over \hbar}(q_1-q_2) - {p_1-p_2 \over \hbar} 
\left( q-{q_1+q_2 \over 2} \right) \Bigg]  \Bigg\},
\label{WCAT0}
\end{eqnarray} 
where $W_{p'q'}(p,q)$ is given by Eq. (\ref{WPQPQ}), does not exist. 
The situation changes during the measurement. 
In the case of a nonselective process, the system is described by a
Wigner function $W(p,q,t)$ which can be exactly evaluated by solving 
Eq. (\ref{PCLWEQ}) with the initial condition (\ref{WCAT0}) \cite{NOTA3}.
The result is 
\begin{eqnarray}
W(p,q,t) = N^2 &\Bigg\{& W_{p_1q_1}(p,q,t)+W_{p_2q_2}(p,q,t) 
+W_{{p_1+p_2\over 2}{q_1+q_2 \over 2}}(p,q,t)
e^{-C_{p_1-p_2q_1-q_2}+\Sigma_{p_1-p_2q_1-q_2}(t)}
\nonumber \\ && \times
2\cos \Bigg[ {p_1+p_2\over 2\hbar}(q_1-q_2) + 
\Upsilon_{p_1-p_2q_1-q_2}(t) 
\left( p-{p_1+p_2 \over 2}e^{-\gamma(t-t')} \right)
\nonumber \\ && 
+\Phi_{p_1-p_2q_1-q_2}(t)  
\left( q-{q_1+q_2 \over 2} -{p_1+p_2 \over 2 m\gamma}
\left( 1 -e^{-\gamma(t-t')} \right) \right) \Bigg]  \Bigg\},
\label{WCAT}
\end{eqnarray} 
where
\begin{eqnarray}
W_{p'q'}(p,q,t) &=& 
{1 \over 2\pi \sqrt{ 4C_{xx}(t)C_{yy}(t) - C_{xy}(t)^2 }}
\nonumber \\ && \times
\exp \Bigg\{ -{C_{xx}(t) \over 4C_{xx}(t)C_{yy}(t) - C_{xy}(t)^2 }
\left[ q-q'-{p'\over m\gamma} \left( 1 -e^{-\gamma(t-t')} \right) \right]^2
\nonumber \\ &&
+{C_{xy}(t) \over 4C_{xx}(t)C_{yy}(t) - C_{xy}(t)^2 }  
\left[ q-q'-{p'\over m\gamma} \left( 1 -e^{-\gamma(t-t')} \right) \right]
\left[ p-p'e^{-\gamma(t-t')} \right]
\nonumber \\ && 
-{C_{yy}(t) \over 4C_{xx}(t)C_{yy}(t) - C_{xy}(t)^2 }  
\left[ p-p'e^{-\gamma(t-t')} \right]^2 \Bigg\},
\label{WT1}
\end{eqnarray}
\begin{eqnarray}
\Sigma_{p'q'}(t) &=&
{ C_{xx}(t) C_{p'q'}^y(t)^2 -C_{xy}(t) C_{p'q'}^x(t) C_{p'q'}^y(t)
+C_{yy}(t) C_{p'q'}^x(t)^2 \over 4C_{xx}(t)C_{yy}(t) - C_{xy}(t)^2 },
\label{GAMMAPQ} 
\end{eqnarray}
\begin{eqnarray}
\Upsilon_{p'q'}(t) &=& 
{2 C_{yy}(t) C_{p'q'}^x(t) - C_{xy}(t) C_{p'q'}^y(t) 
\over 4C_{xx}(t)C_{yy}(t) - C_{xy}(t)^2}, 
\label{UPSILONPQ}
\end{eqnarray}
and
\begin{eqnarray}
\Phi_{p'q'}(t) &=& 
{2 C_{xx}(t) C_{p'q'}^y(t) - C_{xy}(t) C_{p'q'}^x(t) 
\over 4C_{xx}(t)C_{yy}(t) - C_{xy}(t)^2} 
\label{PHIPQ}
\end{eqnarray}
are given in terms of the coefficients
\begin{eqnarray}
C_{xx}(t) &=& \hbar m \gamma \left\{ 
{1\over 2}{\hbar \over m \gamma \sigma_q^2}
\left( {1 \over 4} + {\sigma_{pq}^4 \over \hbar^2} \right) e^{-2\gamma(t-t')} 
+{1\over 2}{k_BT \over \hbar \gamma} \left[ 1-e^{-2\gamma(t-t')}\right] 
\right\},
\\
C_{xy}(t) &=& \hbar \Bigg\{ {\hbar \over m \gamma \sigma_q^2} 
\left( {1\over 4} + {\sigma_{pq}^4 \over \hbar^2} \right)
\left[ 1 - e^{-\gamma(t-t')} \right]e^{-\gamma(t-t')} +
{\sigma_{pq}^2\over \hbar} e^{-\gamma(t-t')}
\nonumber \\ &&
+{k_BT \over \hbar\gamma} \left[ 1-2e^{-\gamma(t-t')}+e^{-2\gamma(t-t')} 
\right] \Bigg\},
\\
C_{yy}(t) &=& {\hbar \over m\gamma} \Bigg\{ {1\over 2}
{m \gamma \sigma_q^2 \over \hbar} 
\left[ 1+ {\sigma_{pq}^2 \over m \gamma \sigma_q^2} 
\left[1-e^{-\gamma(t-t')}\right] \right]^2 + 
{1\over 8} {\hbar \over m\gamma \sigma_q^2} 
\left[1-e^{-\gamma(t-t')}\right]^2
\nonumber \\ &&+{k_BT \over \hbar \gamma}\left[
\gamma(t-t')-{3 \over 2}+2 e^{-\gamma(t-t')}-{1 \over 2}e^{-2\gamma(t-t')}
\right] +{1\over 16} {\hbar \gamma \over k_BT} \gamma (t-t') \Bigg\},
\\
C^x_{p'q'}(t) &=& p' {\sigma_{pq}^2 \over \hbar} 
e^{-\gamma(t-t')} - q' {\hbar \over \sigma_q^2} \left( {1\over 4} +
{\sigma_{pq}^4 \over \hbar^2} \right) e^{-\gamma(t-t')},
\\
C^y_{p'q'}(t) &=& p' {\sigma_{q}^2 \over \hbar}
\left\{ 1 + {\hbar \over m \gamma \sigma_q^2} {\sigma_{pq}^2 \over \hbar}
\left[ 1 - e^{-\gamma(t-t')} \right] \right\}
\nonumber \\ &&
- q' \left\{ {\hbar \over m \gamma \sigma_q^2} \left( {1\over 4} +
{\sigma_{pq}^4 \over \hbar^2} \right) \left[ 1 - e^{-\gamma(t-t')} \right]
+ {\sigma_{pq}^2 \over \hbar} \right\}.
\end{eqnarray}
The indices $x$ and $y$ stand for the Fourier variables conjugated 
to $p$ and $q$, respectively. 
Each term of Eq. (\ref{WCAT}) is localized within a phase space region 
whose size grows with time.
When this size has became much larger than $\sigma_p\sigma_q \sim \hbar$, 
we can write
\begin{equation}
\int dp'dq'~ W(p',q',t) ~W_{p'q'}(p,q) \simeq W(p,q,t)
\end{equation}
and identify the weights $\overline{\delta(p-p(t)) \delta(q-q(t))}$
of Eq. (\ref{LAST}) with  $W(p,q,t)$ in agreement with \cite{HALLIWELL96}.

Let us now turn the classical limit of Eq. (\ref{WCAT}).
First, we note that $W_{p'q'}(p,q,t)$ is the time evolution of the Wigner
function (\ref{WPQPQ}) corresponding to the coherent state $|p'q'\rangle$.
Its classical limit exists and is given by an expression
$W_{p'q'}^{\text{cl}}(p,q,t)$ identical to Eq. (\ref{WT1}) with 
$C_{xx}(t)$, $C_{xy}(t)$, and $C_{yy}(t)$ replaced with 
\begin{eqnarray}
C_{xx}^{\text{cl}}(t) &=& \lim_{\hbar \to 0} C_{xx}(t) =
{1\over 2} mk_BT \left[ 1-e^{-2\gamma(t-t')}\right],
\\
C_{xy}^{\text{cl}}(t) &=& \lim_{\hbar \to 0} C_{xy}(t) =
{k_BT \over \gamma} 
\left[ 1-2e^{-\gamma(t-t')}+e^{-2\gamma(t-t')} \right],
\\
C_{yy}^{\text{cl}}(t) &=& \lim_{\hbar \to 0} C_{yy}(t) =
{k_BT \over m \gamma^2} 
\left[ \gamma(t-t')-{3 \over 2}+2 e^{-\gamma(t-t')}-
{1 \over 2}e^{-2\gamma(t-t')} \right]. 
\end{eqnarray}
The function $W_{p'q'}^{\text{cl}}(p,q,t)$ is the 
phase space probability density obtained by solving the Fokker-Plank
Eq. (\ref{FOKKERPLANK}) with initial condition
$W_{p'q'}^{\text{cl}}(p,q,t')=\delta(p-p')\delta(q-q')$.

Concerning the interference term in Eq. (\ref{WCAT}),  
we have $\Sigma_{p'q'}(t')=C_{p'q'}$, $\Upsilon_{p'q'}(t') = q'/\hbar$, 
and $\Phi_{p'q'}(t') = -p'/\hbar$ so that, as previously noted, 
the $\hbar \to 0$ limit of this term does not exist at $t=t'$ 
due to the undamped oscillation of the cosine. 
On the other hand, for $t>t'$ since $\Sigma_{p'q'}(t) ={\cal O}(\hbar^{-1})$ 
while $C_{p'q'} ={\cal O}(\hbar^{-3/2})$ we have an exponentially 
damping term which allows to obtain 
\begin{equation}
\lim_{\hbar \to 0} W(p,q,t) = 
{1\over 2} \left[ 
W_{p_1q_1}^{\text{cl}}(p,q,t)+W_{p_2q_2}^{\text{cl}}(p,q,t) \right].
\label{WCATCL}
\end{equation}
From a physical point of view, this limit is equivalent to a macroscopic
one in which $|p_1-p_2|/ \sigma_p$ and/or $|q_1-q_2|/ \sigma_q$ 
become infinitely large so that $C_{p_1-p_2q_1-q_2}$ diverges.
In particular, this is obtained by taking the mass $m$ of the particle 
infinitely large.

Finally, we note that due to the condition $\hbar\gamma \ll k_BT$ 
the Wigner function $W_{p'q'}(p,q,t)$ approaches the classical phase space 
probability density $W_{p'q'}^{\text{cl}}(p,q,t)$ on a time scale  
$(\hbar / \gamma k_BT)^{1/2} \ll \gamma^{-1}$ \cite{NOTA4}.
On the other hand, the functions $\Sigma_{p'q'}(t)$, $\Upsilon_{p'q'}(t)$, 
and $\Phi_{p'q'}(t)$ vanish for $t\to \infty$. 
The long time limit of Eq. (\ref{WCAT}), therefore, is
\begin{eqnarray}
W_\infty(p,q,t) = N^2 &\Bigg\{& W_{p_1q_1}^{\text{cl}}(p,q,t)+
W_{p_2q_2}^{\text{cl}}(p,q,t) 
+W_{{p_1+p_2\over 2}{q_1+q_2 \over 2}}^{\text{cl}}(p,q,t)
\nonumber \\ && \times
e^{-C_{p_1-p_2q_1-q_2}}
2\cos \Bigg[ {p_1+p_2\over 2\hbar}(q_1-q_2) \Bigg]  \Bigg\}
\label{WCATLT}
\end{eqnarray} 
and never coincides with the classical limit (\ref{WCATCL}).


\begin{references}
\bibitem{NOTA2} Equations (\ref{SQLO}) and (\ref{SPQLO}) are also the 
leading order terms of Eqs. (\ref{SIGMAQ}) and (\ref{SIGMAPQ}) with respect
to the small parameter $\hbar \gamma/k_BT$. 
They correspond to the particular case $V(x)=0$ and $b=0$ given
in Eq. (3.27) of Ref. \cite{HALLIWELL95B}. 
In that paper the definitions of $\gamma$ and of the Wiener process 
are different from ours and this implies different numerical factors
which, however, cancel out in these simplified expressions of $\sigma_p^2$,
$\sigma_{pq}^2$, and $\sigma_q^2$.
\bibitem{NOTAGAMMA} Note that our definition of $\gamma$ is twice as that
used in \cite{DIOSI93A,DIOSI93B}.
\bibitem{NOTA4} The time scale in question can be defined by the 
conditions  
$C_{xx}^{\text{cl}}(t)/C_{xx}(t)=\case{1}{2}$,
$C_{xy}^{\text{cl}}(t)/C_{xy}(t)=\case{1}{2}$, and 
$C_{yy}^{\text{cl}}(t)/C_{yy}(t)=\case{1}{2}$,
the value of these ratios being 0 at $t=t'$ and monotonically 
approaching 1 for $t\to \infty$.
For $\hbar \gamma /k_BT \ll 1$ and $\gamma (t-t') \ll 1$, each one 
of the above conditions is an equation for the unknown  
$\gamma(t-t')(\hbar \gamma/k_BT)^{-1/2}$ whose solution is 
of the order of unity.
\end{references}
\end{document}